\documentclass[10pt,reqno]{amsart}
\usepackage{amsthm,amsfonts,amsmath,latexsym,amssymb}
\usepackage{amssymb,latexsym,amsfonts,amsmath,amsthm}
\usepackage{hyperref}

\usepackage[pdftex]{graphicx} 
\usepackage{color}
\usepackage{cite}
\def\a{\alpha}
\def\b{\beta}
\def\g{\gamma}
\def\d{\delta}
\def\e{{\rm e}}
\def\eps{\varepsilon}
\def\i{{\rm i}}
\def\s{\sigma}

\def\G{\Gamma}
\def\O{\Omega}
\def\l{\lambda}

\def\p{\partial}

\def\bfe{{\boldsymbol e}}
\def\bfx{{\boldsymbol x}}
\def\bfp{{\boldsymbol p}}
\def\bfv{{\boldsymbol v}}
\def\bfw{{\boldsymbol w}}
\def\bfk{{\boldsymbol k}}
\def\bfn{{\boldsymbol n}}
\def\bfs{{\boldsymbol s}}
\def\bft{{\boldsymbol t}}
\def\bfu{{\boldsymbol u}}
\def\bfq{{\boldsymbol q}}
\def\bfy{{\boldsymbol y}}

\def\C{\mathbb C}
\def\A{{\mathcal A}}
\def\F{{\mathcal F}}
\def\Ha{{\mathcal H}}
\def\Z{\mathbb Z}
\def\R{{\mathbb R}}
\def\MR{\R}
\def\N{\mathbb N}
\def\tr{{\rm tr}\,}
\newcommand{\sgn}{\operatorname{{\mathrm sgn}}}

\theoremstyle{plain}
\newtheorem{thm}{THEOREM}
\newtheorem{lemma}[thm]{LEMMA}
\newtheorem{cl}[thm]{COROLLARY}
\newtheorem{prop}[thm]{PROPOSITION}
\newtheorem{assumption}[thm]{ASSUMPTION}

\theoremstyle{definition}

\theoremstyle{remark}
\newtheorem{remark}[thm]{REMARK}
\newtheorem{remarks}[thm]{REMARKS}

\newcommand{\be}{\begin{equation}}
\newcommand{\ee}{\end{equation}}
\newcommand{\bea}{\begin{eqnarray}}
\newcommand{\eea}{\end{eqnarray}}
\newcommand{\beax}{\begin{eqnarray}}
\newcommand{\eeax}{\end{eqnarray}}

\newcommand{\mfr}[2]{{\textstyle\frac{#1}{#2}}}

\begin{document}

{
\noindent\footnotesize
To appear in \emph{International Mathematics Research Notices} (Oxford)}
\vspace{1.5cm}

\title[``Quadratic" Widom Conjecture]{A Special Case Of A Conjecture
  By Widom With Implications To Fermionic Entanglement Entropy\footnote{
  LMU--ASC 11/09}}

\author[R.C. Helling]{Robert Helling}
\address{Arnold Sommerfeld Center\\
        Ludwig--Maximilians--Universit\"at M\"unchen\\   
        Theresienstra\ss e 37, 80333 M\"unchen}
        \email{helling@atdotde.de}
\author[H.~Leschke]{Hajo Leschke}
\address{Institut f\"ur Theoretische Physik\\
    Universit\"at Erlangen--N\"urnberg\\
    Staudtstra\ss e 7, 91058 Erlangen
    }
    \email{hajo.leschke@physik.uni-erlangen.de} \email{wolfgang.spitzer@physik.uni-erlangen.de}
\author[W.L.~Spitzer]{Wolfgang Spitzer}

\date{9.~April 2010}

\begin{abstract} We prove a special case of a conjecture in asymptotic analysis
by Harold Widom. More precisely, we establish the leading and next--to--leading 
term of a semi--classical expansion of the trace of the square of certain 
integral operators on the Hilbert space $L^2(\R^d)$. As already observed by  
Gioev and Klich, this implies that the bi--partite entanglement entropy of the 
free Fermi gas in its ground state grows at least as fast as the surface area 
of the spatially bounded part times a logarithmic enhancement.
\end{abstract}

\maketitle

\section{Introduction}

In contrast to systems of classical physics, a quantum system
composed of two distinguishable parts may be in a pure
state which is not a product of pure states of its subsystems.
Consequently, if the total system is in such an \emph{entangled} pure state,
the partial state of each subsystem is not pure, in other words, mixed.
Following Einstein, Schr\"odinger, Bell, and others, entanglement may be used to
rule out interpretations of quantum mechanics which are both local and realistic, 
similar to those of classical (statistical) mechanics. More recently,
entanglement has been established as a key concept of quantum communication 
and information theory. For example, quantum teleportation 
and quantum computing heavily rely on it~\cite{Aud,Haya,Nielsen,Stolze}. 

Partially triggered by the latter theories quantifications of entanglement 
(e.g., in terms of entropy) and consequences thereof are at 
present intensively discussed for states of many--particle systems. We refer to 
the reviews by Amico, Fazio, Osterloh, and Vedral~\cite{AFOV}, and by Peschel and 
Eisler~\cite{Peschel}. Here, several interesting results and
conjectures were put forward. However, in most cases a mathematical proof is 
not yet available and one relies on heuristic arguments, approximate 
calculations and/or numerical observations. This is even true for the 
entanglement entropy of the ground states of quantum spin--chains (see Vidal, 
Latorre, Ricco, and Kitaev~\cite{Vidal}) and of a system being as simple 
as the free Fermi gas. Since in the latter system there is no interaction at 
all, a non--trivial entanglement entropy is solely due to
the effective coupling of the particles by the Fermi--Dirac statistics, the 
algebraic statement of Pauli's exclusion principle.

The interest in entanglement entropy was also sparked from quantum field theory,
and in particular by toy models for the Beken\-stein--Hawking entropy of 
black holes~\cite{Beken}. Srednicki~\cite{Sred} and
Bombelli, Koul, Lee, and Sorkin~\cite{BKLS} found numerically that in a 
semi--classical limit which corresponds to scaling the bounded region $\Omega$ 
in $d$--dimensional Euclidean position space $\R^d$ by $R>0$ and taking 
$R\to\infty$, the bi--partite entanglement 
entropy is not a bulk property but scales with the area $R^{d-1}|\partial\O|$ 
of the boundary surface rather than the volume $R^d|\Omega|$. This so--called
\emph{area law} is thought to be generic for field theories with a 
spectral gap above the ground--state energy. See also the more recent works by 
Cramer, Eisert, and Plenio~\cite{CEP} and by Cramer and Eisert~\cite{CE} who 
proved the area--law scaling for harmonic lattice systems. 
It has been suggested that entanglement might be the mechanism 
behind the black--hole entropy. At first, Bekenstein and Hawking
found that black holes behave thermally if one interprets the
surface gravity as temperature and the area of the horizon
as entropy. Especially, there is a ``second law" which states that in
physical processes the total horizon area can never decrease. It is
a major challenge for a quantum theory of gravity to show that this is not
merely an analogy but that the area of the horizon is indeed proportional to
physical entropy and to give a microscopic explanation thereof. In the
framework of string theory this was achieved for extremal black holes by 
Strominger and Vafa~\cite{Strominger} and Maldacena and Strominger~\cite{Maldacena}. 
More generally, it has been argued that the entanglement entropy scales as 
$R^{d-1}$ for $d\ge2$ space dimensions, while for $d=1$ one expects a
\emph{logarithmic} scaling, $\ln R$. In a theory with correlation length 
$\xi<\infty$, heuristic arguments suggest that the entanglement entropy stems 
from correlations across $\partial\Omega$ in a layer of width $\xi>0$
and the absence of long--range correlations is responsible for the area
law. 
However, the area law is observed in conformal field theories for $1+1$
space--time  
dimensions as well, where $\xi=\infty$, see Calabrese and Cardy~\cite{CC,CC2}.

Coming back to simple fermionic systems, Jin and Korepin~\cite{JK} 
showed for the first time that for free fermions on the one--dimensional 
lattice $\Z$ the entanglement entropy for $R\Omega=[-R,R]\cap\Z$ indeed 
scales as $\ln R$, see also Fannes, Haegeman, and Mosonyi~\cite{FHM}.
Wolf~\cite{Wolf} and later Farkas and Zimboras~\cite{Farkas} then proved for 
$d\ge2$ and cubic $\O\subset\Z^d$ a lower bound on the partial particle--number 
variance that scales as $R^{d-1}\ln R$. This, in turn, implies that the 
entanglement entropy grows at least as fast as $R^{d-1}\ln{R}$, thereby 
ruling out an area law. Barthel, Chung and Schollw\"ock~\cite{bart} and 
independently Li, Ding, Yu, Roscilde, and Haas~\cite{Li} provided numerical 
support that the entropy itself scales in the same way up to a numerical
factor (for $d=2$ and $d=3$).

To our knowledge, Gioev and Klich~\cite{GK} were the first to observe 
an intimate connection between the scaling of the entanglement entropy of the 
free--Fermi--gas ground state and an important conjecture in asymptotic analysis by 
Harold Widom~\cite{Widom,Widom1,Widom2}. This ``Widom conjecture"
concerns a two--term asymptotic expansion of the trace, $\tr F(A)$, for a wide 
class of analytic functions $F$ of certain integral operators $A$ on the 
Hilbert space $L^2(\R^d)$, see Equation (\ref{widom}) below. The conjecture may
be understood as a multi--dimensional generalization of Szeg\H{o}'s asymptotics
for Toeplitz determinants and of Slepian's spectral asymptotics in classical information 
theory on the capacity of a communication channel which is band limited in both 
frequency and time. 
In a similar vein, Gioev~\cite{Gioev} established, among other things, for 
the ground state of the free Fermi gas in $\R^d$ and rather general 
$\O\subset \R^d$ with smooth $\p\O$ a lower bound on the partial
particle--number variance that scales as $R^{d-1}\ln{R}$. The main result of 
the present paper establishes an $R^{d-1}\ln{R}$ behavior of that variance itself 
and provides the precise pre--factor in terms of a simple surface integral 
times a numerical constant. Our result is, in fact, a proof of a special case 
of the Widom conjecture for quadratic $F$.


Although we only have a lower bound on the entanglement entropy of the
free--Fermi--gas ground state, we believe, in accordance with a conjecture by 
Gioev and Klich~\cite{GK}, that this bound reflects the correct scaling of the 
entropy itself up to a numerical factor, which is independent of the 
Fermi sea $\G\subset\R^d$ characterizing the ground state, and the region 
$\O\subset\R^d$. As they pointed out, their conjecture actually goes 
beyond the Widom conjecture because the entropy corresponds to a 
non--analytic function $F$ (see our Remark \ref{remark}(iv)). Regardless of the
validity of the Gioev--Klich conjecture, their works~\cite{Gioev,GK} were key 
stimuli to us and apparently also to the authors of \cite{bart,Li}.


The structure of the present paper is as follows: In the next section we
formulate the Widom conjecture. In Section \ref{sec:proof} we prove the Widom 
conjecture for quadratic polynomials $F$. 
Then we proceed in Section \ref{sec:ent} to compile some background material
on fermionic entanglement entropy and apply our result. In Section 
\ref{sec:outlook} we give an outlook of how to possibly prove the Widom 
conjecture for arbitrary polynomials. The paper ends with appendices on the 
method of stationary phase, on the decay properties of certain Fourier 
integrals, and on a simple extension of Roccaforte's estimate 
on the volume of certain self--intersections~\cite{Rocca}.

After having finished the first version of this paper we have learned from 
Alexander Sobolev~\cite{Sobolev} that he has a proof of the Widom conjecture 
for all polynomials, $F$, based on pseudo--differential--operator calculus. 
We are grateful for the explanation of his remarkable achievement prior to 
publication.

\section{The ``quadratic" Widom conjecture}

We start with some notation used throughout the paper. If $d\ge2$, we denote a 
vector $\bfv\in\R^d$ by a boldface letter, and write $v:=|\bfv|:=(\bfv\cdot 
\bfv)^{1/2}$ for its norm. Here, we use a dot to denote the Euclidean scalar
product $\bfv\cdot\bfw$ of two vectors $\bfv,\bfw$ in $\R^d$.
By $A+B:=\{{\boldsymbol a}+{\boldsymbol b}:{\boldsymbol a}\in A,
{\boldsymbol b}\in B\}$ we denote the arithmetic (or Minkowski) sum of a
pair of subsets $A,B\subseteq \R^d$. We also write $A+{\boldsymbol b}
:=A+\{{\boldsymbol b}\}$ for $A\subseteq \R^d$ translated by ${\boldsymbol b}
\in\R^d$ and $R A:=\{R {\boldsymbol a}:{\boldsymbol a}\in A\}$ for
$A\subseteq\R^d$ multiplied by $R\in\R$. 
For a Borel set $\Lambda\subseteq \R^d$ we denote its volume with respect to the
$d$--dimensional Lebesgue measure as $|\Lambda|:=\int_\Lambda d\bfx = \int_{\R^d}
d\bfx\, \chi_\Lambda(\bfx)$, where $\chi_\Lambda$ stands for the indicator
function of $\Lambda$. In particular, if $\Lambda$ is the positive 
half--line, $\Theta:=\chi_{[0,\infty\,[}$ denotes the right--continuous
Heaviside unit--step function. The Hilbert space of complex--valued, Lebesgue 
square--integrable functions $f:\Lambda\to\C$ is denoted as usual by 
$L^2(\Lambda)$. We use the Bachmann--Landau notation of ``little oh'' and 
``big Oh'' in asymptotic (in)equalities in the sense that for real--valued 
functions $f,g$ on $\R$, we write
\begin{itemize}
\item $f(R)\ge g(R) + o(h(R))\quad$  if $\quad\lim_{R\to\infty}
\frac{f(R)-g(R)}{h(R)} \ge 0$;
\item $f(R)= g(R) + o(h(R))\quad$ if $\quad\lim_{R\to\infty}
\frac{f(R)-g(R)}{h(R)} = 0$;
\item $f(R)= g(R) + O(h(R))\quad\!\!$ if $\quad\limsup_{R\to\infty}
\left|\frac{f(R)-g(R)}{h(R)}\right|<\infty$.
\end{itemize}

Next, we formulate our basic assumption.
\begin{assumption} \label{assump} Let $d\ge2$ and $\O\subset\R^d$ and 
$\G\subset\R^d$ be ($C^\infty$--)smooth, compact, $d$--dimensional 
manifolds--with--boundary. The orientation of $\O$ and of its boundary surface 
$\p\O$ is the one induced from $\R^d$, respectively, from the manifold, and 
similarly for $\G$ and $\p\G$. Let $\a$ be a smooth, complex--valued function 
on an open set in $\R^{d}\times\R^d$ containing $\O\times\G$.
\end{assumption}

Note that for such an $\O$ also the difference $\;\O-\O:=\O+(-1)\O\;$ is a smooth, compact 
manifold--with--boundary. For background material in (Riemannian) differential
geometry we refer to the textbooks \cite{BG,Boothby,Thorpe} without further 
notice. 

\medskip
For two sets $\Omega,\Gamma$, and a function $\a$ as described in Assumption 
\ref{assump} we define for each $R>0$ the integral operator 
$A_{R}:L^2(\O)\to L^2(\O)$ by its kernel 
\be \label{kernel}
a_R(\bfx,\bfy):=\left(\frac{R}{2\pi}\right)^d \int_{\G} d\boldsymbol p\,
\e^{\i R(\bfx-\bfy)\cdot\bfp}\,\a(\bfx,\bfp) 
\ee
in the sense that
\be \label{convolution}
(A_{R} f)(\bfx):= \int_{\O} d\boldsymbol y \,a_R(\bfx,\bfy) f(\boldsymbol y)
\,,\quad \bfx\in\O\,,f\in L^2(\O)\,.
\ee
Because of
\be \int_{\O\times\O} d\bfx_1 d\bfx_2 \,\big|a_R(\bfx_1,\bfx_2)\big|^2
\le \left(\frac{R}{2\pi}\right)^{2d}\,|\O|^2\,|\G|^2\,\|\a\|_{\infty,\O,\G}^2
\,,
\ee
where $\|\a\|_{\infty,\O,\G} := \sup\{|\a(\bfx,\bfp)| : (\bfx,\bfp)\in\O
\times\G\} < \infty$, the operator $A_R$ is in the Hilbert--Schmidt class, see
\cite[Theorem VI.23]{Reed}. By \cite[Theorem VI.22(h)]{Reed}, the square
$A_R^2$ (and consequently each natural power $A_R^k, k\ge3$) is then a
trace--class operator.

We recall that $A_R$ can be trivially extended to an operator on $L^2(\R^d)$ by
viewing $L^2(\O)$ as a subspace of $L^2(\R^d)$ and considering
$\widehat{\chi_\O}\, A_R \,\widehat{\chi_\O}$, where the multiplication operator
$\widehat{\chi_\O}$ is the
orthogonal projection from $L^2(\R^d)$ to $L^2(\O)$. The operators
$\widehat{\chi_\O}\, A_R \,\widehat{\chi_\O}$ and $A_R$ have the same non--zero
eigenvalues with the same multiplicities. Therefore, if $F$ is a complex--valued
function with $F(0)=0$ and being analytic on a disc centered at the origin and
with radius strictly larger than $\|\a\|_{\infty,\O,\G}$, then $\tr F(A_R) = 
\tr F(\widehat{\chi_\O}\, A_R \,\widehat{\chi_\O})$. 

In this and similar situations, Widom~\cite{Widom1,Widom2} conjectured
the beautiful two--term asymptotic expansion 
\bea \label{widom}
\tr F(A_{R})&=& \left(\frac{R}{2\pi}\right)^d \,\int_{\O\times\G} 
d\bfx d\bfp \, F(\a(\bfx,\bfp))
\\
&+&\left(\frac{R}{2\pi}\right)^{d-1} \,
  \ln{R}\,\int_{\p\O\times \p\G} 
  d\s({\boldsymbol x}) d\s({\boldsymbol p})\,
  \big|{\boldsymbol n}_{\bfx} \cdot {\boldsymbol n}_{\bfp} \big| \,
  \widetilde{F}\big(\a(\bfx,\bfp)\big) 
  \,+\, o(R^{d-1}\ln{R})\nonumber\,.
\eea
Here, the linear transformation $F\mapsto\widetilde{F}$ is defined by
\be \label{eqU} 
\widetilde{F}(\xi):=\frac1{4\pi^2}\,\int_0^1 
dt\,\frac{F(t\xi)-tF(\xi)}{t(1-t)}\,,\quad\xi\in\R \,,
\ee
$\bfn_\bfx\in\R^d$ and $\bfn_\bfp\in\R^d$ denote the outward unit normal vector 
at $\bfx\in\p\O$, respectively at $\bfp\in\p\G$, and $\s$ is the canonical
($d-1$)--dimensional area measure on the boundary surfaces 
$\p\O$ and $\p\G$. 

Actually, Widom~\cite{Widom2} proved (\ref{widom}) in the case that
$\G$ is a half--space and $\O$ is compact with smooth $\p\O$. 
In~\cite{Rocca}, Roccaforte considered the case 
$\G=\R^d$ and convolution operators arising from a function $\a$ 
not depending on $\bfx$, and whose Fourier transform is decaying 
sufficiently fast. 
Remarkably, he proved a three--term asymptotic expansion $aR^d+bR^{d-1}+cR^{d-2}
+o(R^{d-2})$ of $\tr F(A_{R})$ for certain analytic functions $F$ and
identified the coefficients $a,b$, and $c$ from geometric properties of $\O$.


If $F(t)=t$ and if $A_R$ is a trace--class operator then one simply has,
\be \tr F(A_R) = \tr A_R = \left(\frac{R}{2\pi}\right)^d \,
\int_{\O\times\G} d\bfx d\bfp \, \a(\bfx,\bfp) = \left(\frac{R}{2\pi}\right)^d 
\,\int_{\O\times\G} d\bfx d\bfp \, F(\a(\bfx,\bfp))
\ee
for \emph{all} $R>0$. Our main result is the special case of (\ref{widom}) with 
$F(t)=t^2$ as $R\to\infty$.
\begin{thm}[``Quadratic" Widom Conjecture] \label{thm:qwidom} Under Assumption 
\ref{assump}, the two--term asymptotic expansion 
\bea \label{quadwid}\lefteqn{\tr (A_{R})^2}\\
&=& \left(\frac{R}{2\pi}\right)^d \,\int_{\O\times\G} d\bfx d\bfp \, 
\a(\bfx,\bfp)^2
- \frac1{4\pi^2}\,\left(\frac{R}{2\pi}\right)^{d-1} \, {\ln{R}}\,
\int_{\p\O\times \p\G} d\s({\boldsymbol x}) d\s({\boldsymbol p})\, \nonumber
    \big|{\boldsymbol n}_{\bfx} \cdot {\boldsymbol n}_{\bfp} \big| \,
    \a(\bfx,\bfp)^2
\\&+&o(R^{d-1}\ln{R})\nonumber
\eea
holds as $R\to\infty$.
\end{thm}

\begin{remarks}
\begin{enumerate}
\item[(i)] \label{remark3i}
For dimension $d=1$ and with $\O$ and $\G$ compact intervals, formula
(\ref{quadwid}) remains true as can be seen by an explicit computation. In this
case, the surface integral is simply the sum of the four values taken by 
the function $\a$ at the four corners of the rectangle $\Omega\times\Gamma$.
\item[(ii)]
Our proof relies on the method of stationary phase (see \cite{Stein}) and an 
expression for the volume of the intersection of a set with its 
translate as an integral over the boundary (see Theorem \ref{lemma:volexpans} 
in Appendix~\ref{sect:rocca}). The proof is elementary in the sense that it 
does not rely on tools from pseudo--differential--operator calculus used by
Widom~\cite{Widom2} and recently by Sobolev~\cite{Sobolev} (see the end of the 
Introduction).
\item[(iii)]
In definition (\ref{kernel}) one could evaluate the ``phase--space function" 
(or ``symbol") $\a$
more generally at $(\bfx+\l(\bfy-\bfx),\bfp)$ with $\l\in[0,1]$ instead of 
choosing $\l=0$. The resulting ``$\l$--quantization of $\a$" would then lead 
to an operator $A_{R,\l}$~\cite{Mart}. Here, $1/R$ plays the role of Planck's 
constant. It can be seen from our proof of Theorem \ref{thm:qwidom} 
that the asymptotic behavior of $\tr A_{R,\l}^2$ as $R\to\infty$ has the same 
leading term and next--to--leading term as $\tr A_{R}^2$ for all $\l\in[0,1]$.

\end{enumerate}
\end{remarks}

\section{Proof of Theorem \ref{thm:qwidom}}\label{sec:proof}
The proof consists of two parts. The first part deals with the leading term 
proportional to $R^d$ and an error term of the order $R^{d-1}$. In the second 
part we show how the term proportional to $R^{d-1}\ln{R}$ emerges. 
 
We start out with a simple change of co--ordinates, ${\boldsymbol u}:= \bfx_1,
{\boldsymbol v}:=\bfx_1-\bfx_{2}$, scale ${\boldsymbol v}$ by $1/R$, and hence
write~\cite[p. 524]{Kat}, 
\bea \label{eq8}
\tr (A_{R})^2
   &=& \int_{\O\times\O} d{\boldsymbol x}_1 d{\boldsymbol x}_2 
     \,a_R(\bfx_1,\bfx_{2})\,a_R(\bfx_2,\bfx_{1})
\\
&=&\int_{\O-\O} d{\boldsymbol v} \int_\O d{\boldsymbol u}
   \,a_R(\boldsymbol u,\boldsymbol u-\boldsymbol v)
   \,a_R(\boldsymbol u-\boldsymbol v,\boldsymbol u)
\\
&=&
\left(\frac{R}{4\pi^2}\right)^d \int_{R(\O-\O)} d{\boldsymbol v} 
\int_{\G\times\G} d\bfp d\bfq \,\e^{\i \bfv\cdot(\bfp-\bfq)}
\\
&&\times\int_{\R^d} d\bfu \,\a(\bfu,\bfp) \a(\bfu-\bfv/R,\bfq)\, 
\chi_\O(\bfu)\,\chi_\O(\bfu-\bfv/R)\,.\nonumber
\eea
First we expand $\a(\bfu-\bfv/R,\bfq)$ at $(\bfu,\bfq)$. The error term is
$O(1/R)$. The integral over $\bfu$ is then of the form ($\varepsilon=R^{-1}$)
\be \label{eq10}
\int_{\O\cap (\O+\varepsilon \bfv)} d\bfu\, f(\bfu)
= \int_{\O} d\bfu\,f(\bfu) - \int_{\O\setminus(\O\cap 
(\O+\varepsilon \bfv))} d\bfu\, f(\bfu)
\ee
with $f(\bfu):=\a(\bfu,\bfp) \a(\bfu,\bfq)$. 

Let us define for each $\bfx\in\O$ the function $\g_{\bfx}:\R^d\to\C,\bfv\mapsto
\g_\bfx(\bfv)$ by
\be \label{gen fourier}
    \g_\bfx(\bfv):=(2\pi)^{-d} \int_\G d\bfp\, \a(\bfx,\bfp) \,
    \e^{\i \bfv\cdot\bfp}\,. 
\ee
Then using the uniform decay, $\sup_{\bfx\in\O}|\g_\bfx(\bfv)|\le 
{C}{v^{-\frac{d+1}{2}}}$ (see Lemma~\ref{Gdecay}), and Parseval's identity 
we obtain for the ``leading term",
\be \Big| \int_{R(\O-\O)} d\bfv \,\int_{\G\times\G} d\bfp d\bfq \,
\a(\bfu,\bfp) \,\a(\bfu,\bfq)\,\e^{\i\bfv\cdot(\bfp-\bfq)} - 
(2\pi)^d \int_\G d\bfp\, \a(\bfu,\bfp)^2\Big| \le C/R\,.
\ee

\begin{figure}[h]
  \centering
  \includegraphics[width=0.3\textwidth]{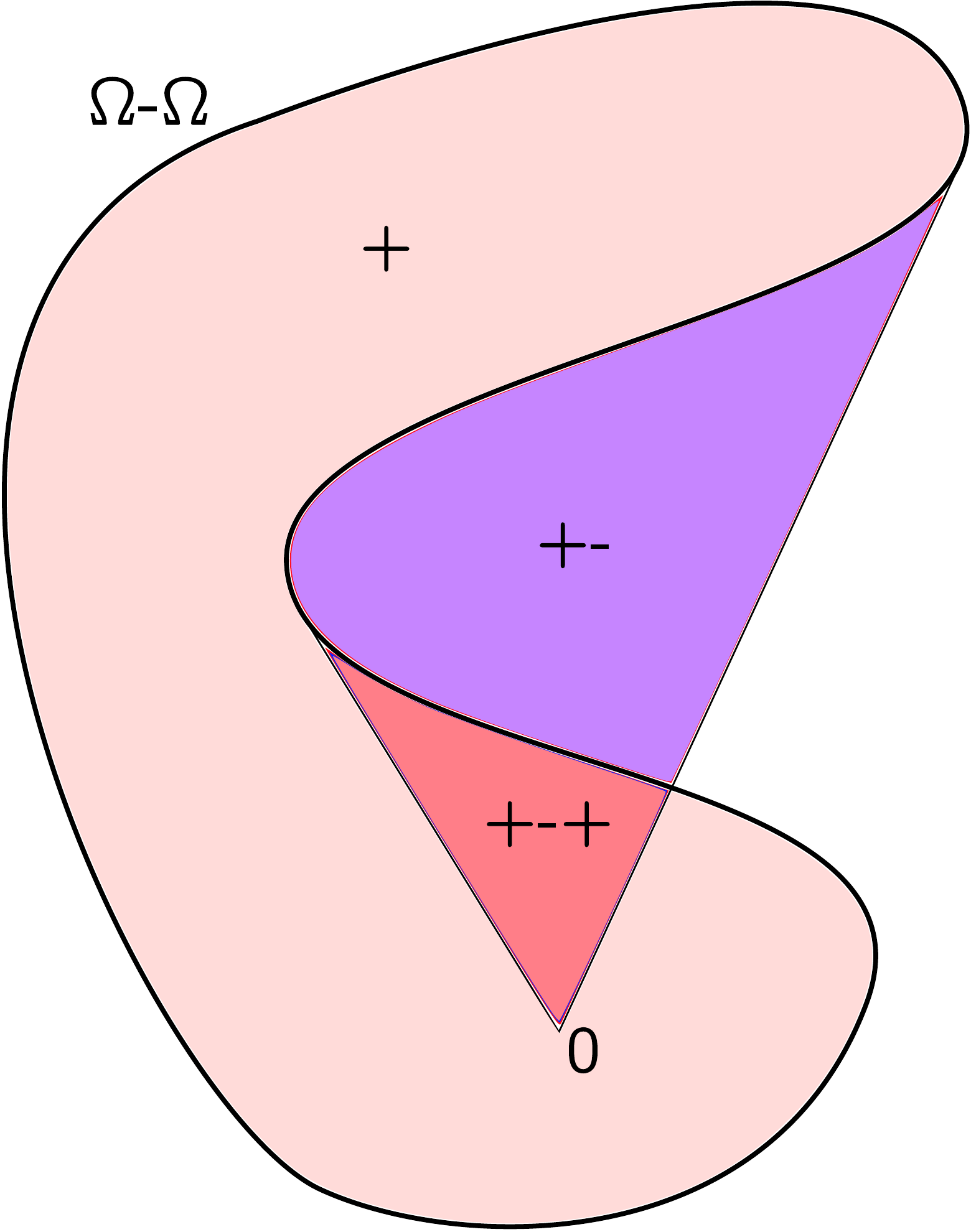}
  \caption{The integration over $\O-\O$.}
  \label{fig:omegaminusomega}
\end{figure}

\medskip
For the second term in equation (\ref{eq10}) we use 
Theorem~\ref{lemma:volexpans} with $\eps=R^{-1}$. Then, after a change of
variables we have to analyze the integral
\be \label{def:I}
I := \left(\frac{R}{2\pi}\right)^{2d}\,\int_{\p\O} d\s(\bfx)
\int_{\O-\O} d\bfv \,\max{(0,\bfv\cdot {\boldsymbol n}_{\bfx})} \,
	\g_\bfx(R\bfv)\g_\bfx(-R\bfv)  \,,
\ee
and a remainder term (proportional to $v^2$) which is easy to deal with
using the decay of $\gamma_\bfx$. Namely,
\begin{equation} R^{2d} \Big|\int_{\O-\O}d\bfv \, v^2 \,\gamma_\bfx(R\bfv)
\gamma_\bfx(-R\bfv) \Big|\le C R^{d-1}
\end{equation} 
for some constant $C$. In order to continue with $I$ from (\ref{def:I}),
it is convenient to write\footnote{Recall that $\Theta$ is the Heaviside 
function.} $\max{(0,\bfv\cdot {\boldsymbol n}_{\bfx})} =  \Theta(\bfv\cdot 
{\bfn}_{\bfx}) \, \bfv\cdot {\bfn}_{\bfx}$. Integrating by parts we get
\bea\label{Gauss}
   (2\pi)^{d} \,\bfv \,\gamma_\bfx(R\bfv) &=& \mfr{1}{\i R} 
   \int_\G d\bfp\, \a(\bfx,\bfp)\,\tfrac{\p}{\p\bfp} 
   \,\e^{\i R\bfv\cdot\bfp}
\\    
&=&\mfr{1}{\i R} \Big(\int_{\p\G} d\s(\bfp)\,{\boldsymbol n}_{\bfp} \,
   \a(\bfx,\bfp)\,\e^{\i R\bfv \cdot\bfp} 
   - \int_{\G} d\bfp\,\big(\tfrac{\p}{\p\bfp}\a(\bfx,\bfp)\big)\,
   \e^{\i R\bfv \cdot\bfp}\Big)
   \,.
\eea 
For the second integral (over $\G$) we may once more integrate by parts and
deduce that it is a term of lower order by another factor of $R^{-1}$. 
Therefore, for some constant $C$, 
\bea \lefteqn{\Big|I + \i \left(\frac{R}{2\pi}\right)^{d}\,R^{-1}\,
\int_{\p\O\times\p\G} \,d\s(\bfx)d\s(\bfp)\,{\bfn}_{\bfx}\cdot {\bfn}_{\bfp}
\,\a(\bfx,\bfp)}\\
&&\times\,R^d\int_{\O-\O} d\bfv\,\Theta(\bfv\cdot {\boldsymbol n}_{\bfx})\,
\g_\bfx(-R\bfv) \,\e^{\i R\bfv\cdot\bfp} \Big|\le C R^{d-1}\,.\nonumber
\eea
The hard part is to analyze the last integral and show that for $\s$--almost
each $\bfp\in\p\G$ one has
\be \label{22}
\Big|R^d\int_{\O-\O} d\bfv\, \Theta(\bfv\cdot {\boldsymbol n}_{\bfx})
\,\gamma_\bfx(-R\bfv) \,\e^{\i R\bfv\cdot\bfp} + (2\pi \i)^{-1} 
\sgn({\boldsymbol n}_{\bfx}\cdot {\boldsymbol n}_{\bfp}) \a(\bfx,\bfp)
\ln{R}\Big| = o(R)\,.
\ee
Here, we need the precise asymptotic expansion (\ref{equ:decay}) for 
$\g_\bfx(\bfv)$ from Lemma~\ref{Gdecay} and the notation employed in its 
proof. For the resulting phase $\bfv\mapsto R(\bfv\cdot\bfp-\bfv\cdot\bfk)$
we are going to apply once more the method of stationary phase. To this end, 
we recall from Lemma~\ref{Gdecay} that each $\bfk$ implicitly depends on $\bfv$
and use generalized polar co--ordinates, $(\rho,\bfw)$, to perform the 
$\bfv$--integration over $\O-\O$. In general, $\bfw\in\partial(\O-\O)$ and 
$\rho\in[0,1]$ are not independent of each other and $\rho$ does not 
necessarily cover the whole interval $[0,1]$. Nevertheless, we may consider the
full interval $[0,1]$ by counting contributions with a negative
sign if at $\bfw\in\partial(\O-\O)$ the boundary is ``inwards'' in the
sense that $\bfw\cdot\bfn_\bfw$ is negative. This is sketched in
Figure \ref{fig:omegaminusomega}. It can be seen that even though
$\rho\bfw$ is not necessarily in $\O-\O$, points outside are counted
with total weight zero while points inside are counted with total weight one.

Let us pick an ortho--normal basis in $\MR^d$ such that the normal vector 
$\bfn_\bfp$ to $\p\Gamma$ at the given point $\bfp\in\p\G$ points in 
the $d$--th direction. Locally around $\bfp$, let $\p\Gamma$ be given by the 
graph of a function $f:U_f\to\R,\bft\mapsto f(\bft)$ with some open $U_f
\subset\R^{d-1}$; in the notation used in the proof of Lemma \ref{Gdecay},
$f=f^{(\bfn_\bfp,m)}$ and $U_f=U_{\bfn_\bfp,m}$ for some $m$. We write
$\bfp=(\bfs,f(\bfs))$ for some $\bfs\in U_f$ and note that, without loss of
generality (by appealing to Sard's Theorem), $f$ is not only critical but has 
an extremum at $\bfs$.

In a similar fashion, we can locally write $\p(\O-\O)$ as the graph of another 
function $g\colon U_g\to\R,\bfu\mapsto g(\bfu)$ with some open
$U_g\subset\R^{d-1}$. 
\emph{We assume for the moment that $\O-\O$ is convex, or put differently that
all boundary points are outwards.} Then, we partition the integration in 
(\ref{22}) into cones $V:=\{(\rho\bfu,\rho g(\bfu)):\rho\in[0,1],\bfu\in U_g\}
\subseteq\O-\O$ with some open set $U_g\subset\R^{d-1}$. Furthermore, instead 
of integrating $\rho$ over $[0,1]$ we may integrate over $[C/R,1]$ for some 
constant $C$ without changing the leading asymptotics of the integral as 
$R\to\infty$. 

The crucial step is to express the condition (see Lemma \ref{new lemma}(i))
$\bfe:=\bfv/v=\sgn{(\bfv\cdot\bfk)}\,\bfn_{\bfk}$ in these new co--ordinates, 
where, without loss of generality, $\bfk=(\bft_\bfk,h(\bft_\bfk))$ with
$h=f^{(\bfe,m')}$ for some $m'$. $\bft_\bfk$ is now thought of as a function of 
$\bfu$. Note that $\bfn_{\bfk}=\sgn{(\bfn_\bfp\cdot\bfn_{\bfk})}(-\mfr{\p}{
\p\bft} h(\bft_\bfk),1)/\sqrt{1+|\mfr{\p}{\p\bft} h(\bft_\bfk)|^2}$ and thus  
\bea\label{crucial equ}
  \frac{(\bfu,g(\bfu))}{\sqrt{\bfu^2+g(\bfu)^2}} &=& \sgn(\bfv\cdot\bfn_{\bfk})
  \sgn{(\bfn_\bfp\cdot\bfn_{\bfk})} \frac{(-\mfr{\p}{\p\bft}
  h(\bft_\bfk),1)}{\sqrt{1+|\mfr{\p}{\p\bft} h(\bft_\bfk)|^2}}\nonumber\\
  &=&\sgn{(g(\bfu))} \,\frac{(-\mfr{\p}{\p\bft}
  h(\bft_\bfk),1)}{\sqrt{1+|\mfr{\p}{\p\bft} h(\bft_\bfk)|^2}}\,.
\eea
Let us proceed with the situation that $(\mathbf 0,g(\mathbf 0))\in V$ and call
this cone $V_0$. As there might be several disjoint graphs of
$f^{(\bfn_\bfp,m)}$, it is notationally simpler to use their union and call the 
corresponding function again $f$. This amounts to setting $h=f$. 

The volume element reads $d\bfv=\rho^{d-1}f(\bfu)$ $d\rho d\bfu$. As $\bfv$ is 
parallel to $\bfn_\bfp=(0,\ldots,0,1)$ at $\bfu=\mathbf 0$ and $V_0$ can be
chosen small enough, we have $\Theta(\bfv\cdot\bfn_\bfx)=\Theta(\sgn(\bfv\cdot
\bfn_\bfp)\bfn_\bfp\cdot\bfn_\bfx)$, and the only contribution to the integral 
is from those $\bfv$ where $g(\bfu)$ has the same sign as the last component 
of $\bfn_{\bfx}$. 
Using the asymptotics from Lemma \ref{Gdecay} we find up to lower--order terms
that (using the abbreviation $\mathcal K_\bfu:=\mathcal K_{(\bfu,g(\bfu))/
\sqrt{\bfu^2+g(\bfu)^2}}$)
\begin{eqnarray}
  \lefteqn{R^d\int_{V_0} d\bfv\, \Theta(\bfv\cdot\bfn_\bfx)
  \g_\bfx(-R\bfv)\,\e^{\i R\bfv\cdot\bfp}}\\
  &=& \i \,(2\pi)^{-\frac{d+1}{2}}\, 
  \int_C^R d\rho\,\rho^{-\frac{d+3}{2}}\int_{U_g} d\bfu\;{\small 
  \frac{g(\bfu)}{(\|\bfu\|^2+g(\bfu)^2)^{(d+1)/4}}}\nonumber\\
  &&\times\,\sum_{\bfk\in\mathcal K_{\bfu}}\,
  {\small \frac{\Theta((\bfu,g(\bfu))\cdot\bfn_\bfx)
  \sgn((\bfu,g(\bfu))\cdot\bfn_{\bfk})}{\sqrt{|\det(f_{ij}(\bft_\bfk))|}}}
  \,\a\big(\bfx,R(\bft_\bfk,f(\bft_\bfk))\big)\, \nonumber
  \\
  &&\times\,\exp{\big[\i R\rho\,(\bfu,g(\bfu))\cdot(\bfp-\bfk)\big]}\,.\nonumber
\end{eqnarray}
Let us write the last $\bfu$--integral in the form $\int_{U_g} d\bfu\,\psi
(\bfu)\,\exp{\big[\i R\rho \phi(\bfu)\big]}$. Then we smoothly extend $\psi$
to a compactly supported complex--valued function $\widetilde{\psi}$ on 
$\R^{d-1}$ and $\phi$ to compactly supported real--valued function 
$\widetilde{\phi}$ on $\R^{d-1}$ in such a way that $\widetilde{\phi}$ does 
not acquire new critical points on the support of $\widetilde{\psi}$ outside 
the support of $\psi$. By Proposition \ref{prop:stein1}, this does not change 
the leading asymptotics of the integral.

Let us investigate now the critical points of the phase function
${\phi}:U_g\subset\R^{d-1}\to\R,\bfu\mapsto(\bfu,g(\bfu))\cdot(\bfp-\bfk(\bfu))$.
Taking derivatives of both sides of equation (\ref{crucial equ}) and evaluating 
at $\bfu=\mathbf 0$ yields (we use the sum convention and sum over indices that
appear twice)
\begin{equation}
  \frac{du_i}{|g(\mathbf 0)|} = -\sgn(g(\mathbf 0))\,
  \frac{\partial^2 f}{\partial t_i\partial t_j}(\bft_\bfk(\mathbf 0)) \,dt_j\,.
\end{equation}
This implies 
\begin{equation}
  \label{eq:partialt}
  \frac{\partial t_j}{\partial u_i}(\mathbf 0) = -\frac{f_{ij}^{-1}
  (\bft_\bfk(\mathbf 0))}{g(\mathbf 0)}\,,
\end{equation}
where $f_{ij}^{-1}$ denotes the matrix inverse of the Hessian of $f$. We are 
now ready to expand the phase to second order in $\bfu$ at $\mathbf 0$:
\begin{eqnarray}
\lefteqn{\bfv\cdot(\bfp-\bfk)/\rho}\\
&=&\bfu\cdot(\bfs-\bft_\bfk(\bfu))+ g(\bfu)\big(f(\bfs)-f(\bft_\bfk(\bfu))\big)
\nonumber\\
&=&\,g(\mathbf 0) \big(f(\bfs) - f(\bft_\bfk(\mathbf 0)\big) \nonumber\\
&& + \, u_i\left(s_i-
  (\bft_\bfk(\mathbf 0))_i
  + \,\frac{\partial g}{\p u_i}(\mathbf 0) \big(f(\bfs)-f(\bft_\bfk(\mathbf
  0))\big)- 
  g(\mathbf 0) \frac{\partial f}{\partial t_j}(\bft_\bfk(\mathbf 0))\, 
  \frac{\p t_j}{\p u_i} (\mathbf 0)\right)\nonumber\\
&&+ \,u_iu_j\left(-\frac{\p t_i}{\p u_j}(\mathbf 0) + 
\frac 12 \frac{\p^2 g}{\p u_i\p u_j} (\mathbf 0)\big(f(\bfs) - f(\bft_\bfk
(\mathbf 0))\big) - \frac 12 \,g(\mathbf 0) \frac{\p^2 f}{\p t_r \p
t_l}(\bft_\bfk(\mathbf 0)) \frac{\p t_r}{\p u_i} \frac{\p t_l}{\p u_j}
(\mathbf 0)\right.
\nonumber\\
&&\qquad\qquad \left. -  \frac{\p g}{\p u_i}(\mathbf 0)  
\frac{\p f}{\p t_r} (\bft_\bfk(\mathbf 0)) \frac{\p t_r}{\p
    u_j}(\mathbf 0)- \frac 12  g(\mathbf 0)  \frac{\p f}{\p t_r}
    (\bft_\bfk(\mathbf 0))
    \frac{\p^2 t_r}{\p u_i\p u_j}(\mathbf 0)\right)\nonumber\\
&& + \,O(u^3)\,.\nonumber
\end{eqnarray}
Using (\ref{eq:partialt}) we obtain 
\bea\lefteqn{
  \bfv\cdot(\bfp-\bfk)/\rho}\\
  &=&  g(\mathbf 0)\big(f(\bfs) - f(\bft_\bfk(\mathbf
  0))\big)  + \bfu\cdot\big(\bfs-\bft_\bfk(\mathbf 0)\big) +\mfr12 u_i u_j
  \Big(\frac {f_{ij}^{-1}(\bft_\bfk(\mathbf 0))}{g(\mathbf 0)} + g_{ij}(\mathbf 0) \big(
  f(\bfs) - f(\bft_\bfk)(\mathbf 0)\big)\Big)\nonumber\\
  && + \,O(u^3)\,.\nonumber
\eea
Then for $V_0$ small enough, the only critical point of the phase function 
${\phi}$ is when $\bft_\bfk(\mathbf 0)=\bfs$ and hence $\bfk = \bfp$. In this
case,
\be \phi(\bfu) = \rho\,\frac {f_{ij}^{-1}(\bft_\bfk(\mathbf 0))}{2g(\mathbf 0)}
 \,u_i u_j \,+ \,O(u^3)\,.
\ee
Next, we apply Proposition \ref{prop:stein2} and conclude that asymptotically 
(up to next--to--leading terms in $1/R$)
\begin{eqnarray}
  &&R^d\int_{V_0} d\bfv\, \Theta(\bfv\cdot\bfn_\bfx)
  \g_\bfx(-R\bfv)\,\e^{\i R\bfv\cdot\bfp} \\
  &&\quad = \i \,(2\pi)^{-\frac{d+1}{2}}
  \int_C^R d\rho\,\rho^{-1} g(\mathbf 0)^{-\frac{d-1}{2}}\,
  \frac{\sgn(\bfn_\bfx\cdot\bfn_\bfp)}{\sqrt{|\det(f_{ij}(\bfs))
  |}}\,\a(\bfx,\bfp)\nonumber
\\
&&\qquad\times\;\exp{\Big[-\frac{\i\pi}4\sgn\big(f_{ij}(\bfs)\big)\Big]}\,
  \int_{\R^{d-1}} d\bfu\,\exp{\Big[-\i\frac{f_{ij}^{-1}
  (\bfs)}{2g(\mathbf 0)}u_iu_j\Big]} \nonumber\\
  &&\quad =(2\pi\i)^{-1}\sgn(\bfn_\bfx\cdot\bfn_\bfp) \,\a(\bfx,\bfp)
  \int_C^R {d\rho}\,\rho^{-1} \nonumber\\
  &&\quad= (2\pi\i)^{-1}\sgn(\bfn_\bfx\cdot\bfn_\bfp)\,\a(\bfx,\bfp)\,
  \ln R\,.\hspace{3cm} \nonumber
\end{eqnarray}
Now we address the other cones in the polar decomposition of $\O-\O$ and show 
that they contribute to lower order. We choose a parametrization $g$ of the
boundary, and cast $\bfw\in\p(\O-\O)$ in the form $\bfw=(\bfu,g(\bfu))$. Let
$\phi:\p(\O-\O)\to\R, \bfw\mapsto \bfw\cdot (\bfp - \bfk(\bfw))$ be the phase  
function in the integral (\ref{22}) in the co--ordinate $\bfw$. We may assume 
that $\bfk(\bfw)\not=\bfp$. Since $\bfw\cdot T_\bfw \bfk = 0$, a point
$\bfw_0$ is a critical point of $\phi$ if and only if $\bfp-\bfk(\bfw_0)$ is
parallel to $\bfn_{\bfw_0}$. In addition, we require that $\phi(\bfw_0)=0$, 
which means that $\bfp - \bfk(\bfw_0)$ is perpendicular to $\bfw_0$. For if 
$\phi(\bfw_0)\not=0$, then the $\rho$--integral $\int_C^R d\rho/\rho\, \exp{[\i 
\rho\phi(\bfw_0)]}$ would be only of the order 1. For a convex hypersurface
$\p(\O-\O)$ these two conditions on $\bfw_0$ cannot be satisfied at once and 
we conclude that only the cone $V_0$ gives the leading contribution. In the 
case of a general hypersurface $\p(\O-\O)$, such cases have to be dismissed
by the fact they form a set of zero surface measure.

Finally, we discuss the case that $\O-\O$ is not convex. By 
the above argument we only need to consider a cone in the direction of
$\bfn_\bfp$. Instead of the 
local parametrization of the boundary surface $\p(\O-\O)$ as the graph of a 
single function $g$ we have, in general, a finite number, $M$, of functions 
$g_n$ defined on open sets $U_{g_n}$ for $1\le n\le M$. We may order them so 
that $U_{g_n}\subset U_{g_{n+1}}$ and that $\mathit{graph}(g_M)$ is the furthest 
part of the boundary surface, and thus the normal vector points outwards. Let 
us define the cones $V_0^{(n)}:=\{(\rho\bfu,\rho g_n(\bfu)):\rho\in[0,1],\bfu
\in U_{g_n}\}$ and let us repeat the above calculation for each of these cone. 
We count their (asymptotic) contributions, namely $(2\pi\i)^{-1}\sgn(\bfn_\bfx\cdot
\bfn_\bfp)\a(\bfx,\bfp)\ln R$, positive/negative if the normal vector at 
$\p(\O-\O)\cap\mathit{graph}(g_n)$ is outwards/inwards. Since $0$ is always in 
the interior of $\O-\O$ and hence $M$ is odd there is only one such term that
survives this summation and we have finished the proof.
\hfill $\Box$

\begin{remark} \label{remark4} 
  We have assumed that the boundaries $\partial\O$ and $\partial\G$
  are smooth. We believe that our proof extends to
  the case of $C^3$--boundaries. Some regularity, however, is needed as
  can be seen from the example of cubes $\O=\G=[-1,1]^d$ with $\a=1$.
  In this case the Fourier transform $\g$, defined in (\ref{Green}), 
  is simply given by the product
  \begin{equation}
  \g(\bfv) = \prod_{i=1}^d \frac{\sin(v_i)}{\pi \,v_i}
  \end{equation}
  with $\bfv = (v_1,\ldots,v_d)$. Hence, the leading decay of $\g(\bfv)$ for large
  $|\bfv|$ is of the form $|\bfv|^{-n}$ with $n\in\{1,\ldots,d\}$ depending on
  the direction $\bfv/|\bfv|$. This is in contrast to the leading decay
  $|\bfv|^{-(d+1)/2}$ in case of a smooth $\p\G$. However, the average decay 
  of the Fourier transform is still of the order $|\bfv|^{-(d+1)/2}$ as was
proved by Brandolini, Hofmann, and Iosevich~\cite{BHI} for convex
sets. In our proof of Theorem (\ref{thm:qwidom}) we critically use the decay
behavior for domains $\Gamma$ fulfilling our Assumption \ref{assump}. It is
obvious that cubes are not covered. However, for the above example of cubes, 
(\ref{quadwid}) can be proved by a direct computation (cf.~Remark 
\ref{remark3i}(i)). 
\end{remark}

\section{Fermionic Entanglement entropy}\label{sec:ent}
We are going to apply Theorem \ref{thm:qwidom} to the ground state of the free 
Fermi gas in the infinitely extended position space $\R^d$. To fix our notation 
and to supply some background material we first consider a slightly more general 
situation.

\subsection{Entanglement entropy of quasi--free fermionic states}
\label{section3.1}

A general system of many fermionic particles with 
separable one--particle Hilbert space $\Ha$ with its scalar product denoted by 
$\langle\cdot,\cdot\rangle$, is described by the (smallest) 
$C^*$--algebra (\!\cite[Theorem 5.2.5]{BR},\cite[Section 4.2]{Ben}) $\A_\Ha$
generated by the unit operator $1$ and the annihilation and
creation operators $a(f)$ and $a^*(g)$ for all $f,g\in\Ha$. These operators 
are bounded and satisfy the usual canonical anti--commutation relations,
\bea \label{CAR} a(f)a^*(g) + a^*(g)a(f) &=& \langle f,g\rangle\,1\,,
\quad f,g\in\Ha\,,\\
a(f)a(g) + a(g)a(f) &=&0\,.  \eea 

A \emph{state} $\rho$ is a linear functional $\rho:\A_\Ha\to\C$ with $\rho(1)=1$
and $\rho(X^*X)\ge0$ for all $X\in\A_\Ha$. A state $\rho$ is called 
\emph{quasi--free} (and \emph{gauge--invariant}) \cite[p.~43]{BR} if there 
exists a self--adjoint operator $D$ on $\Ha$ with ${ 0}\le D\le{ 1}$, such that
\be \label{D}
\rho(a^*(f)a(g)) = \langle g,Df\rangle\,,
\ee 
and, more generally, 
\be\label{wick}
\rho\big(a^*(f_1)\cdots a^*(f_m) a(g_1)\cdots a(g_n)\big) = \left\{
\begin{array}{lcc}
0&\mbox {if } &m\not=n\\
\det\langle g_i,Df_j\rangle &\mbox {if } &m=n\end{array}\right.
\ee
for all finite sets $\{f_1,\ldots,f_m,g_1,\ldots,g_n\}\subset\Ha$. In this
sense a quasi--free $\rho$ is a \emph{generalized Gaussian} state, where $D$ 
plays the role of the covariance.  
We call $D$ the \emph{one--particle density operator} characterizing $\rho$. 
We note that $\rho$ is \emph{pure} if and only if $D$ is a projection, that is, 
$D^2=D$. 

In order to define the (von Neumann) entropy of a quasi--free state we first
introduce the function
\be \label{eta}
\eta(t) := \left\{\begin{array}{lll}0 &\mbox{ if }&t\in\{0,1\}\\
-t\ln{t} - (1-t)\ln{(1-t)}&\mbox{ if } & t\in\;\;]0,1[ \end{array}\right.\,.
\ee 
Now, if $\eta(D)$ is a trace--class operator then the (von Neumann) 
\emph{entropy}, $S(\rho)$, of the quasi--free $\rho$ characterized by
$D$ may be defined as (see~\cite[Equation (6.9)]{Ohya}),
\be \label{Neumann} S(\rho) := \tr \,\eta(D) \,.
\ee
It follows that the entropy of a quasi--free state $\rho$ is zero if and only 
if $\rho$ is pure; this equivalence remains true for non quasi--free states
but we refrain here from defining the entropy for general states.

For a general state $\rho$ and an orthogonal decomposition $\Ha=\Ha_1\oplus
\Ha_2$ into two closed subspaces $\Ha_1$ and $\Ha_2$ we use the isomorphism $\A_\Ha\cong
\A_{\Ha_1}\otimes\A_{\Ha_2}$ to define two \emph{partial} (marginal or reduced) 
states $\rho_1$ and $\rho_2$ on $\A_{\Ha_1}$ and $\A_{\Ha_2}$, respectively, by
\bea \rho_1(X) &:=& \rho(X\otimes {1})\,,\quad X\in\A_{\Ha_1}\,,
\\
     \rho_2(X) &:=& \rho({1}\otimes X)\,,\quad X\in\A_{\Ha_2}\,.
\eea
Then one has the ``triangle'' 
inequality comprised of the Araki--Lieb inequality~\cite{AL} and the 
subadditivity of entropy~\cite[Theorem 6.15]{Ohya},
\be\label{triangle} 
   \big|S(\rho_1) - S(\rho_2)\big| \le S(\rho) \le S(\rho_1) + S(\rho_2)\,.
\ee
Here, the left--hand side is zero by definition if $S(\rho_1)=S(\rho_2)=\infty$.
As a consequence of (\ref{triangle}), the \emph{partial entropies} $S(\rho_1)$ 
and $S(\rho_2)$ are equal if the (total) state $\rho$ is pure.
A simple quantification of the correlations between the subsystems corresponding
to ${\Ha_1}$ and ${\Ha_2}$ in the state $\rho$ of the total system, not 
present in the product state $\rho_1\otimes\rho_2$, is the (bi--partite)
\emph{entanglement entropy},
\be \Delta S(\rho):= S(\rho_1\otimes\rho_2) -  S(\rho) = 
S(\rho_1) + S(\rho_2) - S(\rho)\ge0\,.
\ee
For a pure state $\rho$ this simplifies to \be \label{entangleentrop}
\Delta S(\rho) = 2S(\rho_1) = 2S(\rho_2)\,.
\ee
In words, for a pure state the entanglement entropy is just twice
its partial entropies.

If $\rho$ is quasi--free, then $\rho_1$ and $\rho_2$ are quasi--free, too. More
precisely, if $\rho$ is characterized by $D$ on $\Ha$ as above, then 
$\rho_\ell$ ($\ell\in\{1,2\}$) is characterized by the partial one--particle 
density operator
\be D_\ell := E_\ell DE_\ell\,,
\ee
where $E_\ell:\Ha\to\Ha_\ell$ is the orthogonal projection from $\Ha$ onto 
$\Ha_\ell$. Since $D_1$ (resp.~$D_2$) is the zero--operator on $\Ha_2$ 
(resp.~$\Ha_1$) it is naturally identified with an operator on $\Ha_1$ 
(resp.~$\Ha_2$). By construction, the following identities hold,
\bea \rho_\ell(a^*(f)a(g)) &=& \langle g,D_\ell f\rangle\,,\quad\mbox{etc. (in
analogy to \eqref{wick})} 
\\
     S(\rho_\ell)&=&\tr\,\eta(D_\ell)\,.
\eea

In the special case that the mean of the total number of particles is finite, 
that is, $\tr D<\infty$, then the state $\rho$ is 
given~\cite[Theorem 5.2.14 \& pp.~36--37]{BR} by a density 
operator $W$ on the fermionic Fock space $\F(\Ha)$ over $\Ha$. This positive 
operator of unit trace may be written as
\be\label{eqA}
W=\det(1-D)\,\exp\big[{-\sum_{n,m}\langle f_n,\ln(D^{-1}-1) f_m\rangle \,
a^*(f_n) a(f_m)}\big]\,,
\ee
where $\{f_n\}$ is an arbitrary ortho--normal basis of $\Ha$. Then one 
has~\cite[2.5,11, p.~401]{Thirring4},\cite{Wich}
\be\label{eqB}
S(\rho) = \tr\eta(D) = -\tr W\ln{W} \,,
\ee
which motivates our definition (\ref{Neumann}). We stress that $\tr D<\infty$ 
is not sufficient for $\tr \eta(D)<\infty$ if $\Ha$ has infinite dimension. 
Conversely, the example $D=1$ shows that $\tr\eta(D)=0<\infty$ is possible 
although $\tr D = \infty$.

In the case that (only) $\tr D_1<\infty$, then (at least) $\rho_1$ uniquely 
corresponds to a density operator $W_1$ on $\F(\Ha_1)$ given by a formula 
analogous to (\ref{eqA}). Accordingly, one then has 
\be \label{eq43}
S(\rho_1) = \tr\eta(D_1) = -\tr W_1\ln{W_1}\,.
\ee

Sometimes it is convenient to consider besides the von Neumann entropy also a
more general (but not subadditive) entropy dating back to R\'enyi. 
More precisely, if $\tr D_1<\infty$, we define  
the partial R\'enyi entropy of order $\b$ as (cf.~\cite[Section II.G]{Wehrl})
\be S_\b(\rho_1):=\mfr1{1-\b}\,\ln\tr W_1^\b\,,\quad 
\b\in \,]0,\infty[\,\setminus\,\{1\}\,.
\ee
Note that $S_\b(\rho_1)\ge0$ and $\lim_{\b\to1}S_\b(\rho_1) = S(\rho_1)$. 
Moreover, the Jensen inequality implies the monotonicity,
\be\label{eqE}
(\b-\b')\,\big(S_{\b'}(\rho_1) - S_\b(\rho_1)\big) \ge0\,.
\ee
It may be viewed as a special case of an inequality between (fractional)
absolute moments of a random variable, dating back at least to a work of
Schl\"omilch in 1858, see~\cite[p.~26]{HLP}.
In analogy to (\ref{eq43}) the quasi--free nature of $\rho_1$ implies 
\be\label{eqD} 
S_\b(\rho_1) = \tr \eta_\b(D_1)\,,
\ee
where
\be\label{eq45a}
\eta_\b(t):= \mfr1{1-\b} \ln{\big(t^\b + (1-t)^\b\big)}\,,\quad t\in[0,1]\,.
\ee
For later use we also mention the chain of estimates
\bea\label{eqG}
\lefteqn{2 \,\tr D_1(1-D_1) \le S_2(\rho_1) \le (4\ln 2)\,\tr D_1(1-D_1) 
\le S(\rho_1)}\\
&&\le S_{1/2}(\rho_1) \le 2 \,\tr D_1^{1/2} (1-D_1)^{1/2}
\le 2 \,\tr D_1^{1/2} \,.\nonumber
\eea
The first three estimates follow from $2t(1-t)\le\eta_2(t) \le (4\ln{2})t(1-t) 
\le \eta(t)$ if $t\in[0,1]$. The fourth one is (\ref{eqE}) with
$\b=1/2$ and $\b'\to1$, and the last two follow from $\eta_{1/2}(t)\le 2 t^{1/2}(1-t)^{1/2}\le 2
t^{1/2}$. We now see that $\tr D_1^{1/2}<\infty$ is not only sufficient for $\tr
D_1 <\infty$ but also for $\tr\eta(D_1) <\infty$.

While $\tr D_1$ is physically interpreted as the mean of the number of
particles, the quantity $\tr D_1(1-D_1)$ occurring in (\ref{eqG}) is the 
variance of that number in the quasi--free state $\rho_1$ of the subsystem 
corresponding to $\Ha_1$.

\subsection{Entanglement entropy of the ground state of the free Fermi gas}
Now we consider the special case of a free, spinless Fermi gas in 
$d$--dimensional Euclidean space $\R^d,d\in\N$, at zero
absolute temperature, that is, in its ground state. In the terminology of
Section 4.1 this state $\rho$ is quasi--free and characterized by the Fermi
projection $D=\Theta(\mu-H)$ on $\Ha=L^2(\R^d)$. Here, $H=h(-\i
\frac{\p}{\p\bfx})$ is a translation--invariant one--particle Hamiltonian 
given in terms of a smooth ``dispersion'' function, $h:\R^d\to\R$, on momentum
space, which tends to infinity near infinity and ensures that $H$ is a 
self--adjoint operator on $\Ha$. The prime example is $h(\bfp)=p^2$, 
corresponding to the non--relativistic kinetic energy (in the absense of a 
magnetic field). The real parameter $\mu>\inf\{h(\bfp):\bfp\in\R^d\}$ is the 
Fermi energy. Obviously, one has $\tr D=\infty$ but $S(\rho)=\tr\eta(D)=0<
\infty$ due to $D^2=D$. The Fermi sea corresponding to the Fermi projection 
is given as the lower level set
\be\label{defI}
\G=\{\bfp\in\R^d:h(\bfp) \le \mu\}
\ee
in momentum space. 

In order to study the finite--volume properties of the 
Fermi gas we consider a Borel set $\O\subset\R^d$ with finite volume $|\O|$ 
and thus choose $\Ha_1=L^2(\O)$ and $\Ha_2=L^2(\R^d\setminus\O)$.
Then, according to Section \ref{section3.1}, the partial state
$\rho_1=:\rho_\O$ of that part of the Fermi gas with bounded position
space $\O$ is quasi--free and characterized by
\be\label{defII}
D_1 = \widehat{\chi_\O} \,\Theta(\mu-H)\,\widehat{\chi_\O}=:D_\O\,.
\ee
We may therefore identify $D_\O$ with the operator $A_1$ defined in
(\ref{convolution}) with the function $\a=1$ and $\G$ given by (\ref{defI}).
Moreover, one has (cf.~\cite[p.~524]{Kat} for the calculation of $\tr D_\O$)
\be \label{eq50}
\tr D_\O^2 \le \tr D_\O = (2\pi)^{-d} |\O| |\G| <\infty\,,
\ee
and, by (\ref{eqG}), even
\be \label{defIII}
S(\rho_\O) = \tr \eta(D_\O) \le 2\,\tr D_\O^{1/2} <\infty\,.
\ee
Here, the finiteness of $\tr D_\O^{1/2}$ and hence that of the partial entropy
$S(\rho_{\O})$ of the free Fermi gas in its (pure) ground state $\rho$, was
proved by Gioev and Klich~\cite{GK} by using certain decay properties of 
singular values due to Birman and Solomyak~\cite{BS} (see also Chang and 
Ha~\cite{CH}). We mention in passing that the mean particle \emph{density},
$\tr D_\O/|\O| = |\G|/(2\pi)^d$ is a non--decreasing function of $\mu$.

\medskip
Theorem \ref{thm:qwidom} has the following
\begin{cl}[Lower bound on fermionic entropy] \label{corollary} Suppose $\G$
of (\ref{defI}) and $\O$ satisfy Assumption \ref{assump}. Then, the partial 
entropy $S(\rho_{R\Omega})$ of the free--Fermi--gas ground state
satisfies the asymptotic inequality
\be \label{lower bnd} S(\rho_{R\Omega}) \ge \frac{\ln{2}}{\pi^2}\,
\left(\frac{R}{2\pi}\right)^{d-1} \,\ln{R} \,\int_{\p\O\times \p\G} 
d\s({\boldsymbol x}) d\s({\boldsymbol p})\,
\big|{\boldsymbol n}_{\bfx} \cdot {\boldsymbol n}_{\bfp} \big| 
+\, o(R^{d-1}\ln{R})\,.
\ee
\end{cl}

\begin{remarks}
\begin{enumerate}
\item[(i)] It was already observed by Gioev and Klich~\cite{GK} that a proof 
of (\ref{quadwid}) with the function $\a=1$ would imply (\ref{lower bnd}). 
As mentioned in the Introduction, Gioev~\cite[Inequalities (1.8) \& (1.9)]{Gioev} 
has previously established
a smaller lower bound on $S(\rho_{R\Omega})$ with the same $R^{d-1}
\ln{R}$--scaling.

\item[(ii)] 
An important consequence of the $R^{d-1}\ln{R}$--scaling of the leading term 
in (\ref{lower bnd}) is that it rules out an area law for the entanglement 
entropy in the sense that $\liminf_{R\to\infty}\frac{2 S(\rho_{R\O})}{R^{d-1}} 
= \infty$. The $\mu$--dependence of that term is encoded in the Fermi surface
$\p\G$. 

\item[(iii)] Gioev and Klich~\cite{Gioev,GK} also provided an upper bound on 
$S(\rho_{R\Omega})$ which is, however, larger by an extra factor $\ln{R}$. No 
smaller upper bound is known to us.

\item[(iv)] \label{remark} One may also consider the partial R\'enyi entropies 
$S_\b(\rho_{R\O})$. 
For instance, if $\b=2$, then (\ref{eqG}) gives lower and upper bounds  
on $S_2(\rho_{R\O})$ in terms of the partial particle--number variance
$\tr D_{R\O}(1-D_{R\O})$, which both scale as 
$R^{d-1}\ln{R}$. More generally, by an informal application of the Widom 
conjecture (\ref{widom}) with $\a=1$ to the (non--analytic) function 
$F=\eta_\beta$ from (\ref{eq45a}) and using $\widetilde{\eta}_\b(1) 
= (1+\b)/(24 \b)$ 
it is tempting to conjecture the exact leading asymptotic behavior of the 
partial R\'enyi entropy of order $\b$ to be
\be \label{54}
S_\b(\rho_{R\O}) = \frac{1+\b}{24\b}\,\left(\frac{R}{2\pi}\right)^{d-1} \,
\ln{R}\,\int_{\p\O\times \p\G} d\s({\bfx}) d\s({\bfp})\,
\big|{\bfn}_{\bfx} \cdot {\bfn}_{\bfp} \big| +\, o(R^{d-1}\ln{R}) \,.
\ee
The von Neumann limit $\b\to1$ of (\ref{54}) has already been conjectured by 
Gioev and Klich~\cite{GK} and has stimulated the authors of \cite{bart,Li}.
To our knowledge, the validity of (\ref{54}) is open even for $d=1$ and 
compact intervals $\O$ and $\G$ (cf.~Remark \ref{remark3i}(i)). See, however, 
Jin and Korepin~\cite[Equation (4)]{JK} for non--interacting fermions on the 
one--dimensional lattice $\Z$.
\end{enumerate}
\end{remarks}

\begin{proof}[Proof of Corollary \ref{corollary}] In the (conventional) 
definition (\ref{convolution}) of the operator $A_R$ one keeps $\O$ fixed and 
(effectively) scales $\G$ by $R$. Here, we need to interchange the roles of the  
two sets since physically the ground state of the Fermi gas in $\R^d$, and 
hence its Fermi sea $\G$ is fixed. And one wants to understand the asymptotic 
growth of the entanglement entropy with increasing volume $|\O|$ of the position
space $\O$.
The required interchangeabilty is justified by the fact that the two products 
$QPQ$ and $PQP$ in terms of two arbitrary orthogonal projection operators $Q$ 
and $P$ (on $L^2(\R^d)$) have the same non--zero
eigenvalues with the same multiplicities. This follows from the singular--value
decompositions of $QP$ and $PQ$, see e.g.~\cite[Section 1.2]{Simon}.
Using the third inequality in (\ref{eqG}) for a lower bound, recalling from
(\ref{eq50}) that $\tr A_{R} = \tr D_{R\O} = (\frac{R}{2\pi})^{d} |\Omega| 
|\Gamma|$, and applying Theorem \ref{thm:qwidom} with $\a=1$ finally gives 
(\ref{lower bnd}).

\end{proof}

\section{Outlook}\label{sec:outlook}

Now we show a possible route towards a proof of the Widom conjecture for 
polynomials of arbitrary degree. The reader will have noticed that the 
essential difficulty is already present for the special case $\a=1$, and that 
the extension to general $\a$ is rather straightforward. In what follows we 
will therefore put $\a=1$. Then $\g_{\bfx}(\bfv)$ of (\ref{gen fourier}) 
reduces to the simple Fourier integral,
\be \label{Green}
    \gamma(\bfv) := (2\pi)^{-d} \int_{\G} d{\boldsymbol p} 
    \;\e^{\i \bfv\cdot \bfp} \,,\quad \bfv \in \R^d\,.
\ee
It reproduces itself under convolution, that is, $\g\ast\g=\g$, reflecting the
identity $\chi_\G^2=\chi_\G$.

Proceeding as in equation (\ref{eq8}) we write for $k\in\N$
\be \label{eq48}
\tr (A_{R})^k = \int_{\R^{kd}} \prod_{j=1}^k d\bfx_j\,
\gamma(\bfx_j-\bfx_{j+1}) \chi_{R\O}(\bfx_j) \,,\quad \bfx_{k+1} := \bfx_{1}\,,
\ee
and introduce new co--ordinates $\bfy_0:=\bfx_1,\bfy_1:=\bfx_2-\bfx_1,\ldots,
\bfy_{k-1}:=\bfx_k-\bfx_{k-1}$. Note that $\bfy_0\in R\O,\bfy_1\in R\O-
\bfy_0,\ldots,\bfy_{k-1}\in R\O - \bfy_0 -\ldots-\bfy_{k-2}$. Then 
\bea \label{eq47}\tr (A_{R})^k &=& \int_{\R^{(k-1)d}} d\bfy_1\cdots 
     d\bfy_{k-1}\,\gamma(-\bfy_1)\cdots\gamma(-\bfy_{k-1}) \gamma(\bfy_1+
     \ldots +\bfy_{k-1}) 
\\
&&\times\, 
\int_{\R^d} d\bfy_0\,\chi_{R\O}(\bfy_0)
\chi_{R\O}(\bfy_0+\bfy_1)\cdots\chi_{R\O}(\bfy_0+\ldots+\bfy_{k-1})\,.\nonumber
\eea
For the last integral we write (using Lemma \ref{lemma:volexpans})
\beax\lefteqn{\int_{\R^d} d\bfy_0\,\chi_{R\O}(\bfy_0)\chi_{R\O}(\bfy_0+\bfy_1)
\cdots\chi_{R\O}(\bfy_0+\ldots + \bfy_{k-1})= }\\
&=&R^d |\O| - R^d \big|\O\setminus(\O\cap (\O-\bfy_1/R)\cap \ldots\cap 
(\O-\bfy_1/R-\ldots-\bfy_{k-1}/R))\big|\nonumber
\\
&=&\Big[R^d |\O| - R^{d-1} \int_{\p\O} d\s(\bfx) \,
\max(0,\bfy_1\cdot \bfn_{\bfx},\ldots,(\bfy_1+\ldots+\bfy_{k-1})\cdot 
\bfn_{\bfx})\\
&&\;+ \,O(R^{d-2})\Big]\,\times\, \nonumber
\chi_{R(\O-\O)}(\bfy_1)\cdots\chi_{R(\O-\O)}(\bfy_1+\ldots+\bfy_{k-1})\,.
\eeax
For the leading term in (\ref{eq48}) we obtain $\left(\frac{R}{2\pi}\right)^d
|\O| |\G|$ by the same argument as for $k=2$ with an error $O(R^{d-1})$; recall
that $\g\ast\cdots\ast\g=\g$.

Since in (\ref{eq47}) the function $(\bfy_1,\ldots,\bfy_{k-1})\mapsto\g(-\bfy_1)
\cdots\g(-\bfy_{k-1})\g(\bfy_1+\ldots +\bfy_{k-1})$ is symmetric we only need
to consider the symmetric part of the remaining function, namely of
\begin{align}\nonumber\lefteqn{(\bfy_1,\ldots,\bfy_{k-1})\mapsto
\max(0,\bfy_1\cdot \bfn_{\bfx},\ldots,(\bfy_1+\ldots+\bfy_{k-1})\cdot
\bfn_{\bfx})}\\&&\qquad\times\,\chi_{R(\O-\O)}(\bfy_1)\cdots
\chi_{R(\O-\O)}(\bfy_1+\ldots+\bfy_{k-1})\,.
\label{eq55}
\end{align}
The maximum function by itself can be easily symmetrized by the following quite
surprising combinatorial lemma.
\begin{lemma} \label{comb lemma} Let $a_1,\ldots,a_n$ be real numbers. Then
\be \label{combident}
\sum_{\s} \max(0,a_{\s1},a_{\s1}+a_{\s2},\ldots,a_{\s1}+\ldots+a_{\s n})
=\sum_\s \sum_{\ell=1}^n\frac1\ell \max(0,a_{\s1}+\ldots+a_{\s \ell})\,,
\ee
where on both sides the summation $\sum_\s$ runs over the $n!$ permutations of 
$\{1,\ldots,n\}\subset\N$.
\end{lemma}
The lemma was formulated and used in this version by 
Widom~\cite[pp.~171,174]{Widom}. Under the same assumptions, 
Kac~\cite[Theorem 4.2]{Kac} presents a proof (due to F.~Dyson) that
\be \label{Kac}
\sum_{\s} \max(0,a_{\s1},a_{\s1}+a_{\s2},\ldots,a_{\s1}+\ldots+a_{\s n})
=\sum_\s a_{\s1} \sum_{k=1}^n \Theta(a_{\s1}+\ldots+a_{\s k})\,,
\ee
where $\Theta$ is (as above) the Heaviside function. It can be easily shown 
that the right--hand sides of (\ref{combident}) and (\ref{Kac}) are equal and 
hence the combinatorial lemma is proved. 

By the transformation (\ref{eqU}) we obtain for the power function $F(t)=t^k$ 
that $4\pi^2 \widetilde{F}(1) = -\sum_{\ell=1}^{k-1} \frac{1}{\ell}$ which 
fits the right--hand side of (\ref{combident}).

Now we come to the next--to--leading term in (\ref{eq48}), and consider for 
$1\le \ell\le k-1$
\bea \lefteqn{\int_{\R^{(k-1)d}} d\bfy_1\cdots d\bfy_{k-1}\,\g(-\bfy_1)\cdots
\g(-\bfy_{k-1}) \g(\bfy_1+\ldots +\bfy_{k-1})}
\\&&\times \, \chi_{R(\O-\O}(\bfy_1)\cdots
\chi_{R(\O-\O}(\bfy_1+\ldots+\bfy_{k-1}) \,\max(0,(\bfy_1+\ldots+\bfy_\ell)
\cdot \bfn_\bfx)\,.\nonumber
\eea
Then, to leading order, we perform the integration with respect to all 
variables except $\bfv:=\bfy_1+\ldots+\bfy_\ell$. This leaves us with the 
familiar term
\be \int_{R(\O-\O)} d\bfv\, \big|\g(\bfv)\big|^2\,\max(0,\bfv\cdot 
\bfn_\bfx)
\ee
that yields the logarithmic correction term which we know from the $k=2$ 
calculation. 

To complete the proof, one has to show that the error resulting from only
symmetrizing the maximum function but not the product of indicator functions 
in (\ref{eq55}) is of lower order as $R\to\infty$.

\begin{appendix}

\section{Method of Stationary Phase}


We are going to cite two propositions on the method of stationary phase 
that will be used in this paper. To begin with, let us recall that a smooth 
real--valued function $\phi$ on $\R^{d-1}$ has a critical point at
$\bft_0\in\R^{d-1}$ if $\p\phi(\bft)/\p\bft\big|_{\bft=\bft_0} = \boldsymbol 0$.
Such a point is called non--degenerate if the determinant $\det{\phi_{ij}(\bft)}$
of the Hessian $\phi_{ij}(\bft):=\p^2\phi(\bft)/\p t_it_j$ of $\phi$
is non--zero at $\bft=\bft_0$. By $\sgn \phi_{ij}(\bft)$ we denote the number 
of strictly positive minus the number of strictly negative eigenvalues of this 
Hessian at $\bft\in\R^{d-1}$.

\begin{prop} \label{prop:stein1} Let $r$ be a smooth complex--valued and let 
$\phi$ be a smooth real--valued function on $\R^{d-1}$. Moreover, let $r$ have 
a compact support \emph{not} containing a critical point of $\phi$. Then
\be \int_{\R^{d-1}} d\bft\,r(\bft) \,\e^{\i R\phi(\bft)} = O(R^{-N})
\ee
as $R\to\infty$ for any $N\in\N$.
\end{prop}
For a proof see \cite[Chapter VIII, Section 2, Proposition 4]{Stein} or
\cite[Theorem 7.7.1]{Horm}. The second result (see \cite[Chapter VIII, 
Section 2, Proposition 6]{Stein} or \cite[Theorem 7.7.5]{Horm}) 
deals with the asymptotics of the integral in case the phase $\phi$ has a 
non--degenerate critical point.

\begin{prop} \label{prop:stein2}
Let $r$ be a smooth complex--valued and let $\phi$ be smooth real--valued 
function on $\R^{d-1}$. Suppose that $\phi$ \emph{has} a non--degenerate 
critical point at 
$\bft_0$. If $r$ is supported in a sufficiently small neighborhood of 
$\bft_0$ (so that there is no other critical points in its support), then 
there exists a sequence $(z_j)_{j\in\N_0}$ of complex numbers such that
\be \e^{-\i R\phi(\bft_0)} \int_{\R^{d-1}} d\bft\,r(\bft)\, \e^{\i R\phi(\bft)} 
= R^{-(d-1)/2}\left(\sum_{j=0}^{N-1} z_j R^{-j} +O(R^{-N})\right)
\ee
as $R\to\infty$ for any $N\in\N$. In particular, $z_0=r(\bft_0) 
(2\pi)^{(d-1)/2} \big|\det\phi_{ij}(\bft_0)\big|^{-1/2} \e^{\,\i \frac{\pi}{4}
\sgn \phi_{ij}(\bft_0)}$. 
\end{prop}

In the following $K(\bfp)$ denotes the Gauss--Kronecker curvature of $\p\G$ 
at $\bfp\in\p\G$, and $\mathrm{sign}(\bfp)$ is the number of strictly positive 
minus the number of strictly negative eigenvalues of the second fundamental 
form of $\p\G$ at $\bfp\in\p\G$.

\begin{lemma}\label{new lemma}
Let $\G\subset\R^d$ be a smooth, compact, $d$--dimensional
manifold--with--boundary. Then there exists a subset $E$ of the
$(d-1)$--dimensional unit sphere $\mathbb S^{d-1}\subset\R^d$ of full Haar 
measure such that for each $\boldsymbol e\in E$ there exists a non--empty and 
finite set $\mathcal K_\bfe\subset\p\G$ such that for all $\bfk\in \mathcal 
K_\bfe$
\begin{enumerate}
\item[(i)] $\boldsymbol e\cdot \bfn_{\bfk}\in\{-1,1\}$,
\item[(ii)] $K(\bfk)\not=0$.
\end{enumerate}
In other words, $\mathcal K_\bfe$ is the set of non--degenerate critical points
of the mapping $\p\G\to\R, \bfp\mapsto \bfp\cdot\bfe$.
\end{lemma}

As we noted in Remark \ref{remark4} for the cube $\Gamma = [-1,1]^d$, 
smoothness of $\p\G$ cannot be relaxed to piecewise smoothness without 
jeopardizing the non--emptiness and/or finiteness of the sets $\mathcal K_\bfe$. 

\begin{proof} First we recall the definition of Gauss's spherical mapping
$G:\p\G\to\mathbb S^{d-1}$, $\bfp \mapsto G(\bfp):=\bfn_\bfp$, and that the
curvature $K(\bfp)$ is given by the Jacobian determinant of $G$ evaluated at
$\bfp\in\p\G$. Then we define $E$ as 
$E:=G(\{\bfp\in\p\G:K(\bfp)
\not=0\})$. By compactness we have $G(\p\G)=\mathbb S^{d-1}$ and that the 
pre--image $\mathcal K_\bfe:=G^{-1}(\{-\bfe,\bfe\})$ is a non--empty and finite set for 
each $\bfe\in E$. By Sard's Theorem the complement $\mathbb S^{d-1}\setminus 
E$ is of Haar measure 0.
\end{proof}

{\bf Choice of co--ordinates:} From now on we assume that for fixed $\bfe\in E$
the hypersurface $\p\G$ in $\R^d$ is locally given by graphs of certain smooth
functions $f^{(\bfe,m)}:U_{\bfe,m}\to\R$ defined on some open sets 
$U_{\bfe,m}\subseteq\R^{d-1}$ indexed by some $m\in\{1,\ldots,M\}$ for 
some finite $M\in\N$ ($M$ can be chosen independent of $\bfe$). To be more
specific, let us assume for a moment that $\bfe=(0,\ldots,0,1)$. Then, up to a
permutation of the $d$ co--ordinates $\bft=(t_1,\ldots,t_{d-1})$ and 
$f^{(\bfe,m)}(\bft)$, we have $\mathit{graph} f^{(\bfe,m)}:=\{(\bft,
f^{(\bfe,m)}(\bft))\in\R^{d-1}\times\R:\bft\in U_{\bfe,m}\}$ and
$\bigcup_{m=1}^M \mathit{graph} f^{(\bfe,m)} = \p\G$. In addition, we 
may assume that each $\mathit{graph} f^{(\bfe,m)}\subseteq\p\G$ is small 
enough so that it contains at most one $\bfk\in\mathcal K_\bfe$ (see Lemma 
\ref{new lemma}) and each $\bfk\in \mathcal K_\bfe$ is contained in only one of
these graphs. More precisely, for every $\bfk\in \mathcal K_\bfe$ 
there exists a unique $m_\bfk\in\{1,\ldots,M\}$ such that $\bfk\in 
\mathit{graph}f^{(\bfe,m_\bfk)}$ and $\bfk=(\bft_\bfk,f^{(\bfe,m_\bfk)}
(\bft_\bfk))$ for some $\bft_\bfk\in U_{\bfe,m_\bfk}$. Note that such a point
$\bft_\bfk$ is a critical point of $f^{(\bfe,m_\bfk)}$, that is, 
$\frac{\p f^{(\bfe,m_\bfk)}(\bft)}{\p\bft}\Big|_{\bft=\bft_\bfk} = 
\boldsymbol 0$. Furthermore, the curvature of $\p\G$ at $\bfk\in \mathcal 
K_\bfe$ becomes the determinant of the Hessian of $f^{(\bfe,m_\bfk)}$ at 
$\bft_\bfk$ (that is, $K(\bfk) = \det{(f_{ij}^{(\bfe,m_\bfk)}(\bft_\bfk))}
$), and that $\mathrm{sign}(\bfk)= \sgn{(f_{ij}^{(\bfe,m_\bfk)}(\bft_\bfk))}
$. If $\bfe=\mathcal R(0,\ldots,0,1)$ for a suitable rotation $\mathcal R$ 
then we simply rotate the graphs of $f^{(\bfe,m)}$ by this $\mathcal R$.


\begin{lemma}[Decay of the function $\g_{\bfx}$] \label{Gdecay}
Let $\G\subset \R^d,\mathcal K_\bfe\subset\p\G$, and $E\subseteq\mathbb S^{d-1}$ 
be as in Lemma \ref{new lemma}.
Finally, let $\g_{\bfx}(\bfv)$ be defined by (\ref{gen fourier}). Then one has 
for large $v>0$ the asymptotic formula
\begin{equation}
\label{equ:decay} \gamma_\bfx(\bfv) = - \i\,(2\pi v)^{-\frac{d+1}2}
\sum_{\bfk\in \mathcal K_{\bfv/v}}
\frac{\sgn(\bfv\cdot\bfn_{\bfk})}{\sqrt{|K(\bfk)|}} \,
\a(\bfx,\bfk)\,\e^{\i\bfv\cdot\bfk +\frac{\i\pi}4 \mathrm{sign}(\bfk)}\,
\Big(1 +O(v^{-1})\Big) 
\end{equation}
for all $\bfv/v\in E$. 
The remainder term $O(v^{-1})$ is independent of $\bfx\in\O$. 
\end{lemma}

Formula (\ref{equ:decay}) is a slight variant of a standard result
that can be found, for example, in \cite[Theorem 7.7.14]{Horm}. For this
identification we note that $\sgn(\bfv\cdot\bfn_{\bfk})\,\e^{\frac{\i\pi}4 
\mathrm{sign}(\bfk)} = \e^{\frac{\i\pi}4 \sigma(\bfk)}$, where ``H\"ormander's
index'' $\sigma(\bfk)$ denotes the number of centers of curvature at $\bfk$ in
the direction $\bfv/v$ minus the number of centers of curvature at $\bfk$ in
the direction $-\bfv/v$. Nevertheless, since we are using these co--ordinates
in the proof of Theorem \ref{thm:qwidom}, we provide a proof based on 
Propositions \ref{prop:stein1} and \ref{prop:stein2}.
\begin{proof} As in (\ref{Gauss}) we use integration by parts to obtain 
\begin{eqnarray}
(2\pi)^{d} \g_\bfx(\bfv) &=& \frac{\bfv}{\i v^2}\cdot\int_\Gamma 
d\bfp\,\a(\bfx,\bfp)\,\mfr{\p}{\p\bfp}\,\e^{\i\bfv\cdot\bfp}
\\
&=& \frac{\bfv}{\i v^2}\cdot\Big(\int_{\p\Gamma} d\s(\bfp)\, \bfn_{\bfp}\,
\a(\bfx,\bfp)\,\e^{\i \bfv\cdot\bfp} - \int_\G d\bfp\,
\big(\mfr{\p}{\p\bfp}\a(\bfx,\bfp)\big)\,\e^{\i\bfv\cdot\bfp}\Big)
\,.\nonumber
\end{eqnarray}  
For the second integral in the last equation we may repeat the same 
integration--by--parts procedure with $\frac{\bfv}{v^2} \cdot \p\a(\bfx,\bfp)/
\p\bfp$ instead of $\a(\bfx,\bfp)$. This results in a term of the 
order $v^{-\frac{d-3}2}$ but with the same phase as the 
leading term. Therefore the leading term of $\g_\bfx(\bfv)$ as $v\to\infty$ 
stems from the first integral, which we consider in what follows.

Now, let $\bfe:=\bfv/v$ in E be fixed. Moreover, let $(\psi_\l)_\l$ be a finite
$C^\infty$--partition of unity which is subordinate to the covering 
$(\mathit{graph}f^{(\bfe,m)})_m$ of $\p\G$ in the sense that for each
$\l$, $\mathrm{supp}(\psi_\l)\subset\mathit{graph}f^{(\bfe,m(\l))}$
for some uniquely determined $m(\l)\in\{1,\ldots, M\}$. In addition, if 
$\mathcal K_\bfe\cap\mathrm{supp}(\psi_\l)=:\{\bfk(\l)\}$ is non--empty for 
some $\l$, then we require that $\psi_\l(\bfp)=1$ for all $\bfp$ in a 
neighborhood of this $\bfk(\l)$.


\begin{figure}[h]
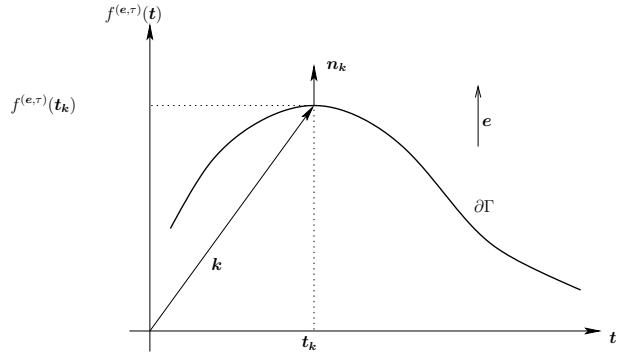

  \centering
  \resizebox{8cm}{!}{\input Gammacoords} 
  \caption{Co--ordinates for the $\bft$--integration.}
  \label{fig:Gammacoords}
\end{figure}

In our co--ordinates we therefore get
\be\label{68}
\bfe\cdot\int_{\p\G} d\s(\bfp) \,\bfn_\bfp \,\a(\bfx,\bfp)\, 
\e^{\i \bfv\cdot\bfp}=\sum_\l \int_{U_{\bfe,m(\l)}}d\bft 
\, r_\l(\bft)\,\e^{\i v \phi_\l(\bft)}\,,
\ee
with certain smooth functions $r_\l$ and $\phi_\l$. In case $\bfe=(0,\ldots,0,
1),\mathit{graph} f^{(\bfe,m)}:=\{(\bft,f^{(\bfe,m)}(\bft))\in
\R^{d-1}\times\R:\bft\in U_{\bfe,m}\}$ as above (see also Figure
\ref{fig:Gammacoords}), and abbreviating $\bfp^{(\l)}(\bft):=
(\bft,f^{(\bfe,m(\l))}(\bft))$ with $\bft\in U_{\bfe,m(\l)}$, we 
explicitly have
\beax r_\l(\bft)&=&\sgn{\big(\bfe\cdot 
\bfn_{\bfp^{(\l)}(\bft)}\big)}\,\bfe\cdot \big(-\p f^{(\bfe,m(\l))}
(\bft)/\p\bft,1\big) \,\psi_\l(\bfp^{(\l)}(\bft))\,\a(\bfx,\bfp^{(\l)}
(\bft))\,,
\\
\phi_\l(\bft)&=&\bfe\cdot \bfp^{(\l)}(\bft)\,.
\eeax 
Formula (\ref{68}) follows from the three facts
\begin{enumerate}
\item[(i)] $\sum_\l\psi_\l(\bfp)=1$, 
\item[(ii)] $d\s(\bfp)=d\bft\sqrt{1+|\p f^{(\bfe,
m(\l))}(\bft)/\p\bft|^2}$ for the area measure on $\mathit{graph}f^{(\bfe,
m(\l))}$, and 
\item[(iii)] $\bfn_\bfp=\sgn(\bfe\cdot\bfn_{\bfp^{(\l)}(\bft)})
\big(-\p f^{(\bfe,m(\l))}(\bft)/\p\bft,1\big)/\sqrt{1+|\p f^{(\bfe,m(\l))}
(\bft)/\p\bft|^2}$ for the unit normal vector at $\bfp\in\mathit{graph}f^{
(\bfe,m(\l))}$.
\end{enumerate}
We note that the signum function in $r_\l$ takes either the value $1$ or $-1$ on
the whole of $U_{\bfe,m(\l)}$.

Next, we want to replace the domain of integration $U_{\bfe,m(\l)}$ on the
right--hand side of (\ref{68}) by $\R^{d-1}$ without changing the value of the
integral. Since $\psi_\l$ has compact support in $\mathit{graph} f^{(\bfe,m)}$,
we smoothly extend $r_\l$ to $\R^{d-1}$ simply by setting $r_\l(\bft):=0$ if 
$\bft\not\in U_{\bfe,m(\l)}$. The phase function $\phi_\l$ is smoothly 
extended by Urysohn's Lemma.

Now, we split the sum in (\ref{68}) into a sum over those $\l$ such that
$\mathcal K_\bfe\cap\mathrm{supp}(\psi_\l)=\emptyset$ and those for which this
intersection is non--empty; in fact, it contains then only a single point,
$\bfk(\l)$. Thus we have 
\bea\lefteqn{\frac{\bfv}v\cdot
\int_{\p\Gamma} d\s(\bfk)\,\bfn_{\bfk}\,
\a(\bfx,\bfk)\,\e^{\i \bfv\cdot\bfk}}\\
& = &
  \sum_{\l:\mathcal K_\bfe\cap\mathrm{supp}(\psi_\l)=\emptyset}
  \int_{\R^{d-1}} d\bft\,r_\l(\bft)\,\e^{\i v \phi_\l(\bft)} 
  + 
  \sum_{\l:\mathcal K_\bfe\cap\mathrm{supp}(\psi_\l)=\{\bfk(\l)\}} 
  \int_{\R^{d-1}} d\bft\,r_\l(\bft)\,\e^{\i v \phi_\l(\bft)}
  \,.\nonumber
\eea
In the first sum we get the decay $v^{-N}$ for any $N$ according to 
Proposition \ref{prop:stein1}. In the second sum, $\phi_\l$ is expanded to 
second order around its critical point $\bft_{\bfk(\l)}$. By Proposition 
\ref{prop:stein2} we therefore arrive at (\ref{equ:decay}).
By compactness of $\O$ we may choose the remainder term $O(v^{-1})$ to be
independent of $\bfx\in\O$. 
\end{proof}

\section{Roccaforte's estimate on the volume of 
self--intersections}\label{sect:rocca}

In \cite[Theorem 2.1]{Rocca}, Roccaforte proved a theorem which is (by one 
order of $\eps$ below) more precise than what we need here. See also a previous 
version by Widom~\cite[Lemma 2 \& 2']{Widom}). But Roccaforte's proof also 
allows for the inclusion of a smooth function in the integrand. For the
convenience of the reader we present his proof almost literally and do not 
claim any originality. Note, however, that the derivative of $f$ effects the 
correction of the order $\eps^2$ but this is not needed here. 

\begin{thm}[Roccaforte]\label{lemma:volexpans} Let $\O\subset\R^d$ be a compact
set with $C^2$--boundary $\p\O$, $\bfv_1,\ldots,$ $\bfv_n\in\R^d$, $\varepsilon>0$, 
and $\O_{\bfv_1,\ldots,\bfv_n} := \O\cap(\O+\bfv_1)\cap\cdots\cap(\O+\bfv_n)$. 
Let $f$ be a $C^1$--function defined on $\O$. Then there exists a constant 
$C$ depending on $\O$ and (the supremum of the derivative of) $f$ so that
\be \Big|\int_{\O\setminus \Omega_{\varepsilon\bfv_1,\ldots,\varepsilon
\bfv_n}} d\bfx\, f(\bfx) + \varepsilon\int_{\p\O} d\s({\boldsymbol x})
\,f(\bfx)\,\max_{1\le k\le n}{(0,\bfv_k\cdot{\boldsymbol n}_\bfx)}\Big| 
\le C \,\varepsilon^2 \,,
\ee
where $\bfn_\bfx$ is the unit outward normal at $\bfx\in\p\O$.
\end{thm}

\begin{figure}[h]
  \centering
  \includegraphics[width=0.4\textwidth]{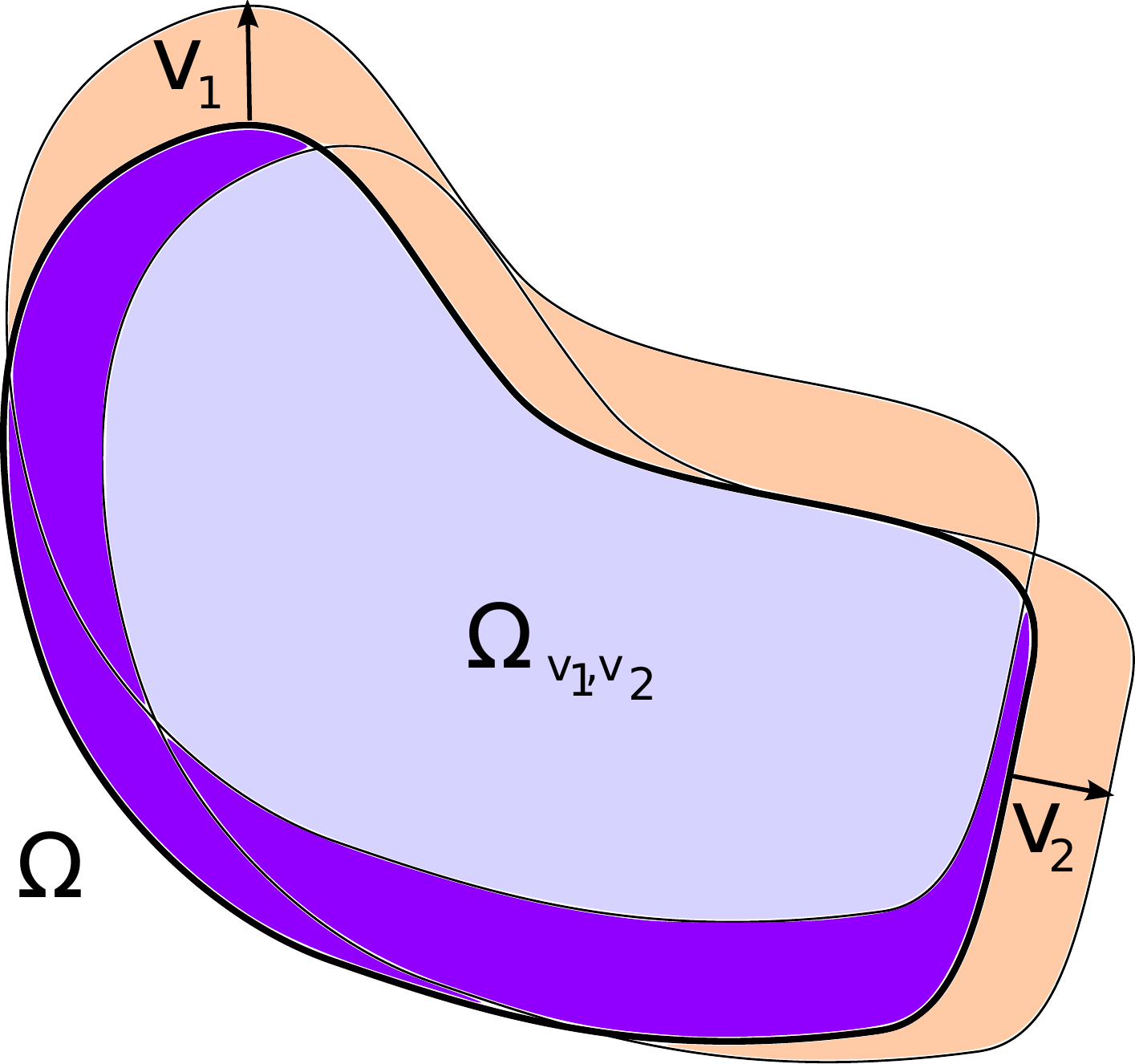}
  \caption{The integration region $\Omega\setminus\Omega_\eps$ is
    marked in dark purple.}
  \label{fig:roccaforte}
\end{figure}


\begin{proof} Let $\O_\eps:=\Omega_{\varepsilon\bfv_1,\ldots,\varepsilon
\bfv_n}$. Since $\p\O$ is compact there exists an $\eps_0$ and an 
$\eps_0$--tubular neighborhood $N_{\eps_0}$ of $\p\O$ such that each $\bfx\in 
N_{\eps_0}$ can be written uniquely as $\bfx = \bar{\bfx} - s\bfn_{\bar{\bfx}}$,
where $\bar{\bfx}\in\p\O$, $\bfn_{\bar{\bfx}}$ is the unit outward normal 
vector at $\bfx$, and $|s|<\eps_0$. If $\eps$ is small enough then 
$\O\setminus\O_\eps\subset N_{\eps_0}$.

Let $\{U_j,\psi_j\}$ be a finite atlas of co--ordinate neighborhoods covering
$\p\O$. Let $V_j:=\{\bfx\in N_{\eps_0}: \mbox{ if }
\bfx = \bar{\bfx} - s\bfn_{\bar{\bfx}}, \mbox{ then }\bar{\bfx}\in U_j\}$.
Define $\phi_j:V_j\to\R^{d-1}\times\R$ as follows: if $\bfx = \bar{\bfx} - s
\bfn_{\bar{\bfx}}\in V_j$ and $\psi_j(\bar{\bfx})=\bar{\bfu}\in\R^{d-1}$, then 
$\phi_j(\bfx):=\bar{\bfu} + s\bfn$, where $\bfn:=(0,\ldots,0,1)$ is the unit
vector in $\R^{d-1}\times\R$ normal to $\R^{d-1}$. By the compactness of $\p\O$
there exist open sets $N_j\subset V_j$ such that the $N_j$ are an open cover of
$N_{\eps_0}$ and the distance from $N_j$ to the complement of $V_j$ is, for all $j$,
greater than some $\eps_1$. If $\eps$ is chosen such that $\max\{|\eps \bfv_k|,
1\le k\le n\}<\eps_1$ then, for all $j,k,\bfx\in N_j$ implies $\bfx-\eps 
\bfv_k\in V_j$. Let $W_j:=N_j\cap\p\O\subset U_j$.

Let $\{\rho_j\}$ be a partition of unity subordinate to the cover $\bigcup_j
W_j$ of $\p\O$. Each $\rho_j$ extends to a function $\widetilde{\rho}_j$ on
$N_j\cap \O\setminus\O_\eps$ by defining $\widetilde{\rho}_j(\bar{\bfx} - 
s\bfn_{\bar{\bfx}}):=\rho_j(\bar{\bfx})$. It now suffices to prove
\be\label{2.1}
\int_{N_j\cap\O\setminus\O_\eps} d\bfx \,f(\bfx) \widetilde{\rho}_j(\bfx) +
\eps\int_{W_j} d\s(\bar{\bfx})\,f(\bar{\bfx}) \rho_j(\bar{\bfx})\,
\max_{1\le k\le n}{(0,\bfv_k\cdot\bfn_{\bar{\bfx}})} = O(\eps^2)\,.
\ee
In what follows the index $j$ will be dropped. From the construction of $\phi$
it follows that for $\bfy\in V$, $\bfy\in\O\cap V$ if and only if $\phi(\bfy)
\cdot\bfn=s\ge0$. Hence for $\bfx\in N$, $\bfx\in N\cap \O_\eps$ if and only if
$s\ge0$ and, for all $1\le k\le n$, $\phi(\bfx-\eps \bfv_k)\cdot\bfn \ge0$. 
By Taylor's Theorem $\bfx\in N\cap\O_\eps$ if and only if $s\ge0$ and, for all 
$k$,
\be\label{2.2}
\phi(\bfx)\cdot\bfn -\eps \,(D_\bfx\phi)(\bfv_k)\cdot\bfn + 
R(\eps) \ge0\,,
\ee
where $R(\eps)=O(\eps^2)$, the estimate being uniform over $\bfx$ since
$\phi\in C^2(V)$; $D_\bfx\phi$ denotes the derivative of $\phi$ at $\bfx$
with matrix elements $(D_\bfx\phi)_{ij} = \frac{\p\phi_i}{\p x_j}(\bfx)$.

Next we show that for any $\bfv\in\R^d$ and all $\bfx = \bar{\bfx} - 
s\bfn_{\bar{\bfx}}\in N$, 
\be \label{lemma2.2}
(D_\bfx\phi)(\bfv)\cdot\bfn = \bfv\cdot\bfn_{\bar{\bfx}}\,.
\ee
To see this, let $\bfv_t:=\bfv-(\bfv\cdot\bfn_{\bar{\bfx}})
\bfn_{\bar{\bfx}}$ be the component of $\bfv$ that is tangent to $\p\O$ at
$\bfx$. Writing
\be (D_\bfx\phi)(\bfv)\cdot\bfn = (D_\bfx\phi)(\bfv_t)\cdot\bfn + 
(\bfv\cdot \bfn_{\bar{\bfx}}) \, (D_\bfx\phi)(\bfn_{\bar{\bfx}})\cdot\bfn
\ee
it suffices to prove
\be\label{2.7}
(D_\bfx\phi)(\bfv_t)\cdot\bfn=0\quad\mbox{ and }\quad 
(D_\bfx\phi)(\bfn_{\bar{\bfx}}) = \bfn \,.
\ee
For each fixed $s_0$ with $|s_0|<\eps_0$ the map $\phi^{-1}:(\bar{\bfu},s_0)
\mapsto \bar{\bfx}(\bar{\bfu}) - s_0\bfn_{\bar{\bfx}}
(\bar{\bfu})$ describes the hypersurface $W-s_0\bfn_{\bar{\bfx}}$. The vectors
$\p(\bar{\bfx} - s_0\bfn_{\bar{\bfx}})/\p u_i(\bar{\bfu}_0)$ form a basis for
the tangent space to $W-s_0\bfn_{\bar{\bfx}}$ at $\bar{\bfx}(\bar{\bfu}_0) - 
s_0\bfn_{\bar{\bfx}}(\bar{\bfu}_0)$. Thus, the derivative $D_{(\bar{\bfu}_0,
s_0)}\phi^{-1}$ sends vectors $(\bar{\bfu},0)$ to vectors tangent to 
$W-s_0\bfn_{\bar{\bfx}}$ at $\bar{\bfx}(\bar{\bfu}_0) - s_0\bfn_{\bar{\bfx}}
(\bar{\bfu}_0)$ and sends $\bfn$ to $\bfn_{\bar{\bfx}}$. Hence it suffices to 
show that for any $\bfx=\bar{\bfx} -s\bfn_{\bar{\bfx}}$, the tangent space 
$T_\bfx(W-s\bfn_{\bar{\bfx}}) = T_{\bar{\bfx}}(\p\O)$. But
$$ \bfn_{\bar{\bfx}} \cdot \frac{\p}{\p u_i}(\bar{\bfx}-s\bfn_{\bar{\bfx}}) =
-s \bfn_{\bar{\bfx}} \cdot \frac{\p}{\p u_i}(\bfn_{\bar{\bfx}}) = 
-\frac{s}2\,\frac{\p}{\p u_i}(\bfn_{\bar{\bfx}}\cdot \bfn_{\bar{\bfx}}) = 0\,.
$$
So, $T_\bfx(W-s\bfn_{\bar{\bfx}})$ is orthogonal to $\bfn_{\bar{\bfx}}$ and
hence the same as the tangent space $T_{\bar{\bfx}}(\p\O)$.

From (\ref{2.2}) and (\ref{lemma2.2}), $\bfx\in N\cap\O\setminus\O_\eps$ if and only
if
\be s\ge0 \quad\mbox{and, for some } 1\le k\le n,\quad
s-\eps \bfv_k \cdot \bfn_{\bar{\bfx}} +R(\eps) <0\,.
\ee
Let 
\[ S_\eps:=\{\bfx\in N: s\ge 0 \mbox{ and for all } 1\le k\le n, 
s-\eps \bfv_k \cdot \bfn_{\bar{\bfx}} +R(\eps) \ge 0\}\,,
\]
and for real $\delta$,
\[I_\delta:=\{\bfx\in N: s\ge 0 \mbox{ and for all } 1\le k\le n, 
s-\eps \bfv_k \cdot \bfn_{\bar{\bfx}} \ge \delta\eps^2\}\,.
\]
Then we have 
\be\label{2.4}
N\cap \O\setminus\O_\eps = N\cap \O\setminus S_\eps\,,
\ee
and the volume of the symmetric difference,
\be\label{lemma2.3}
\big|S_\eps\bigtriangleup I_0\big| = O(\eps^2)\,.
\ee
(\ref{2.4}) is obvious. To prove (\ref{lemma2.3}) note first that since 
$R(\eps)=O(\eps^2)$ there is for each $\delta>0$ an $\eps_\delta>0$ so that
\be\label{2.11}
\eps<\eps_\d \quad\mbox{ implies }\quad I_\d \subset S_\d\subset I_{-\d}\,.
\ee
Next,
\be\label{2.12}
\eps<\eps_\d \quad\mbox{ implies }\quad S_\eps\bigtriangleup I_0 \subset 
I_{-\d}\setminus I_\d\,.
\ee
For, if $\bfx\in S_\eps$ but $\bfx\not\in I_0$ we have by (\ref{2.11}) that
$\bfx\in I_{-\d}$ and $\bfx\not\in I_0$ implies $\bfx\not\in I_\d$. Similarly,
if $\bfx\not\in S_\eps$ but $\bfx\in I_0$ implies $x\in I_{-\d}\setminus I_\d$.
From (\ref{2.12}) it then follows that for $\eps<\eps_\d$,
$$ \big|S_\eps\bigtriangleup I_0\big| \le \big|I_{-\d}\setminus I_\d\big|
=\int_{\phi(I_{-\d}\setminus I_\d)} d\bar{\bfu} ds\,\big|\det 
D_{(\bar{\bfu},s)}\phi^{-1}\big|\,.
$$
But
\be
\phi(I_{-\d}\setminus I_\d) =\big\{\phi(\bfx),\bfx\in N: 
s\ge0, -\d\eps^2 + \eps \max_{1\le k\le n}(\bfv_k\cdot\bfn_{\bar{\bfx}}) \le 
s\le \d\eps^2 + \eps \max_{1\le k\le n} (\bfv_k\cdot\bfn_{\bar{\bfx}})\big\} 
\,.
\ee
Thus the above integral is
$$\le 2\d\eps^2 \sup_{(\bar{\bfu},s)\in\phi(N)} \big|\det 
D_{(\bar{\bfu},s)}\phi^{-1}\big|\,\int_{\phi(\p\O\cap I_{-\d}\setminus I_\d)} 
d\bar{\bfu} \le 2\d\eps^2 M\,,
$$
where $M:=\sup\{|\det D_{(\bar{\bfu},s)}\phi^{-1}|:(\bar{\bfu},s)\in\phi(N)\} 
\,|\phi(W)|$ and $|\phi(W)|$ is the $(d-1)$--dimensional Lebesgue volume of 
$W\subset\R^{d-1}$; by the compactness of $\O$ one can guarantee $M<\infty$. 
This shows (\ref{lemma2.3}).

This allows us to replace $N\cap\O\setminus\O_\eps$ by $N\cap\O\setminus I_0$
in (\ref{2.1}). Changing variables we obtain,
\be\label{2.5}
\int_{\O\cap N\setminus I_0} d\bfx\, (f\widetilde{\rho})(\bfx) = 
\int_{\phi(\O\cap N\setminus I_0)} d\bar{\bfu} ds\,(f\widetilde{\rho})\circ
\phi^{-1}(\bar{\bfu},s)\,\big|\det D_{(\bar{\bfu},s)}\phi^{-1}\big|\,.
\ee
Now we expand $(f\widetilde{\rho})\circ\phi^{-1}$ and $|\det D\phi^{-1}|$ at $s=0$
to first order. Then the last integral equals
\be 
\int_{\phi(\O\cap N\setminus I_0)} d\bar{\bfu} ds\,\big[(f\widetilde{\rho})\circ
\phi^{-1}(\bar{\bfu},0)\,|\det D_{(\bar{\bfu},0)}\phi^{-1}| + O(s)\big]\,,
\ee
where by the definitions of $\phi$ and $I_0$
\be\label{2.6}
\phi(\O\cap N\setminus I_0) = \big\{\phi(\bfx), \bfx\in N:
0\le s\le\eps\max_{1\le k\le n}{(0,\bfv_k\cdot\bfn_{\bar{\bfx}})}\big\}\,.
\ee
Integrating with respect to $s$ yields
\be \eps\int_{\phi(W)} d\bar{\bfu}\,(f\rho)\circ\phi^{-1}(\bar{\bfu},0) \,
|\det D_{(\bar{\bfu},0)}\phi^{-1}|\,\max_{1\le k\le n}{(0,\bfv_k\cdot
\bfn_{\bar{\bfx}})} + O(\eps^2)\,, 
\ee
which proves our statement by another change of variables.
\end{proof}

\end{appendix}

\noindent{\bf Acknowledgment:}  R.H. and W.S. are grateful to the support of 
Jacobs University Bremen, where this work was started. We also thank Urs
Frauenfelder for discussions.

\end{document}